\documentclass[12pt]{article}

\usepackage{amscd}
\usepackage{amsfonts}
\usepackage{amsmath}
\usepackage{amsthm}
\usepackage{amssymb}
\newtheorem{theorem}{Theorem}[]
\newtheorem{lemma}[theorem]{Lemma}

\newtheorem{proposition}[theorem]{Proposition}
\newtheorem{corollary}[theorem]{Corollary}

\theoremstyle{definition}
\newtheorem{definition}[theorem]{Definition}

\newtheorem{remark}[theorem]{Remark}

\newtheorem{example}[theorem]{Example}
\newlength{\fuyasu}
\renewcommand\P{{\mathcal Y}}

\newcommand\M{\overline{M}}
\renewcommand\H{{\mathcal H}}
\newcommand\Hz{{\displaystyle \mathop{\H}^{\circ}}}
\newcommand\R{{\mathcal R}{\mathcal C}}

\newcommand\Z{\mathbb{Z}}
\newcommand\Zn{\Z_{\ge0}}

\newcommand{\fil}{
\setlength{\unitlength}{5mm}
\begin{picture}(1,1)(0,0.3)
\put(0,0){\line(1,0){1}}
\put(0,1){\line(1,0){1}}
\put(0,0){\line(0,1){1}}
\put(1,0){\line(0,1){1}}
\put(0.5,0.5){\circle*{0.7}}
\end{picture}
}
\newcommand{\emp}{
\setlength{\unitlength}{5mm}
\begin{picture}(1,1)(0,0.3)
\put(0,0){\line(1,0){1}}
\put(0,1){\line(1,0){1}}
\put(0,0){\line(0,1){1}}
\put(1,0){\line(0,1){1}}
\end{picture}
}
\newcommand{\filfil}{
\setlength{\unitlength}{5mm}
\begin{picture}(2,1)(0,0.3)
\put(0,0){\line(1,0){2}}
\put(0,1){\line(1,0){2}}
\multiput(0,0)(1,0){3}{\line(0,1){1}}
\put(0.5,0.5){\circle*{0.7}}
\put(1.5,0.5){\circle*{0.7}}
\end{picture}
}
\newcommand{\filemp}{
\setlength{\unitlength}{5mm}
\begin{picture}(2,1)(0,0.3)
\put(0,0){\line(1,0){2}}
\put(0,1){\line(1,0){2}}
\multiput(0,0)(1,0){3}{\line(0,1){1}}
\put(0.5,0.5){\circle*{0.7}}
\end{picture}
}
\newcommand{\empfil}{
\setlength{\unitlength}{5mm}
\begin{picture}(2,1)(0,0.3)
\put(0,0){\line(1,0){2}}
\put(0,1){\line(1,0){2}}
\multiput(0,0)(1,0){3}{\line(0,1){1}}
\put(1.5,0.5){\circle*{0.7}}
\end{picture}
}
\newcommand{\empemp}{
\setlength{\unitlength}{5mm}
\begin{picture}(2,1)(0,0.3)
\put(0,0){\line(1,0){2}}
\put(0,1){\line(1,0){2}}
\multiput(0,0)(1,0){3}{\line(0,1){1}}
\end{picture}
}
\newcommand{\boxarray}{
\put(0.5,1){\line(1,0){24}}
\put(0.5,2){\line(1,0){24}}
\multiput(1,1)(1,0){24}{\line(0,1){1}}
\put(-0.5,0){\makebox(1,1){$\cdots$}}
\put(0.5,0){\makebox(1,1){$0$}}
\put(1.5,0){\makebox(1,1){$1$}}
\put(2.5,0){\makebox(1,1){$2$}}
\put(3.5,0){\makebox(1,1){$3$}}
\put(4.5,0){\makebox(1,1){$4$}}
\put(5.5,0){\makebox(1,1){$5$}}
\put(6.5,0){\makebox(1,1){$6$}}
\put(7.5,0){\makebox(1,1){$7$}}
\put(8.5,0){\makebox(1,1){$8$}}
\put(9.5,0){\makebox(1,1){$9$}}
\put(10.5,0){\makebox(1,1){$10$}}
\put(11.5,0){\makebox(1,1){$11$}}
\put(12.5,0){\makebox(1,1){$12$}}
\put(13.5,0){\makebox(1,1){$13$}}
\put(14.5,0){\makebox(1,1){$14$}}
\put(15.5,0){\makebox(1,1){$15$}}
\put(16.5,0){\makebox(1,1){$16$}}
\put(17.5,0){\makebox(1,1){$17$}}
\put(18.5,0){\makebox(1,1){$18$}}
\put(19.5,0){\makebox(1,1){$19$}}
\put(20.5,0){\makebox(1,1){$20$}}
\put(21.5,0){\makebox(1,1){$21$}}
\put(22.5,0){\makebox(1,1){$22$}}
\put(23.5,0){\makebox(1,1){$23$}}
\put(24.5,0){\makebox(1,1){$\cdots$}}
}
\newcommand{\youngdg}{
\put(0,0){\line(1,0){1}}
\put(0,1){\line(1,0){1}}
\put(0,2){\line(1,0){2}}
\put(0,3){\line(1,0){5}}
\put(0,4){\line(1,0){5}}
\put(0,0){\line(0,1){4}}
\put(1,0){\line(0,1){4}}
\put(2,2){\line(0,1){2}}
\put(3,3){\line(0,1){1}}
\put(4,3){\line(0,1){1}}
\put(5,3){\line(0,1){1}}
}
\title{Inverse scattering method for \\
a soliton cellular automaton}

\author{Taichiro Takagi\\
\normalsize
\em Department of Applied Physics, National Defense Academy,\\
\normalsize
\em Kanagawa 239-8686, Japan}

\date{}
\begin{document}
\maketitle

\begin{abstract}
A set of action-angle variables for a soliton cellular automaton
is obtained.
It is identified with the rigged configuration, a
well-known object in Bethe ansatz.
Regarding it as the set of scattering data
an inverse scattering method to solve initial value problems
of this automaton is presented.
By considering partition functions for this
system a new interpretation of a fermionic character
formula is obtained.
\end{abstract}

% \clearpage
% 
\section{Introduction}\label{sec:1}
\subsection{Action-angle variables for the
box-ball system}
The box-ball system is a well-known example of soliton
cellular automata \cite{T,TS}.
Here we show its simplest version.
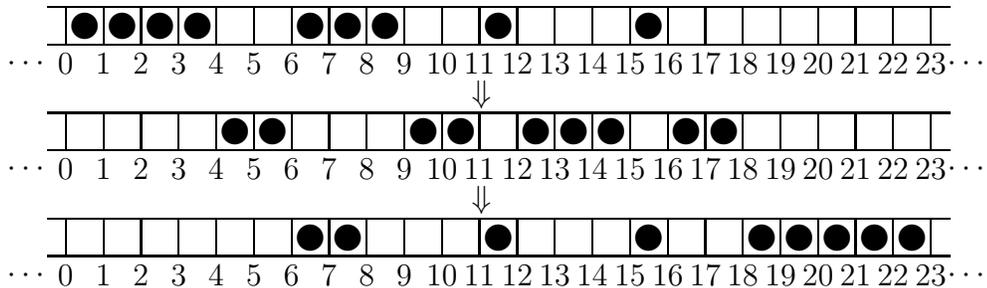
\begin{align*}
&
\setlength{\unitlength}{5mm}
\begin{picture}(23,1)(0,0.7)
\boxarray
\put(1.5,1.5){\circle*{0.7}}
\put(2.5,1.5){\circle*{0.7}}
\put(3.5,1.5){\circle*{0.7}}
\put(4.5,1.5){\circle*{0.7}}
\put(7.5,1.5){\circle*{0.7}}
\put(8.5,1.5){\circle*{0.7}}
\put(9.5,1.5){\circle*{0.7}}
\put(12.5,1.5){\circle*{0.7}}
\put(16.5,1.5){\circle*{0.7}}
\end{picture}
\\
&\qquad \qquad \qquad \qquad \qquad \qquad \qquad
\Downarrow\\
&
\setlength{\unitlength}{5mm}
\begin{picture}(23,1)(0,0.7)
\boxarray
\put(5.5,1.5){\circle*{0.7}}
\put(6.5,1.5){\circle*{0.7}}
\put(10.5,1.5){\circle*{0.7}}
\put(11.5,1.5){\circle*{0.7}}
\put(13.5,1.5){\circle*{0.7}}
\put(14.5,1.5){\circle*{0.7}}
\put(15.5,1.5){\circle*{0.7}}
\put(17.5,1.5){\circle*{0.7}}
\put(18.5,1.5){\circle*{0.7}}
\end{picture}
\\
&\qquad \qquad \qquad \qquad 
\qquad \qquad \qquad
\Downarrow
\\
&
\setlength{\unitlength}{5mm}
\begin{picture}(23,1)(0,0.7)
\boxarray
\put(7.5,1.5){\circle*{0.7}}
\put(8.5,1.5){\circle*{0.7}}
\put(12.5,1.5){\circle*{0.7}}
\put(16.5,1.5){\circle*{0.7}}
\put(19.5,1.5){\circle*{0.7}}
\put(20.5,1.5){\circle*{0.7}}
\put(21.5,1.5){\circle*{0.7}}
\put(22.5,1.5){\circle*{0.7}}
\put(23.5,1.5){\circle*{0.7}}
\end{picture}
\end{align*}
The arrows show the time evolution process of the
automaton whose precise
rule will be given in the main text 
(See Section~\ref{sec:2}, also for the convention of the
position coordinate).
This system and its generalization to accommodate
several kinds of balls
are known as integrable systems
by a certain limiting procedure on discrete KdV/Toda
equations \cite{TNS,TTMS}
or by explicit constructions of their
conserved quantities
\cite{FOY,TTS}.
Recently
Kuniba, Okado, Yamada and the author presented
a conjecture \cite{KOTY} that a set of 
action-angle variables for this system was
given by the {\em rigged configuration},
an object known in Bethe ansatz \cite{KKR}.
In the simplest case
a rigged configuration 
is a Young diagram with a set of integers
{\em rigged} to its rows.
For instance
according to the above example the associated 
rigged configuration evolves as
\begin{displaymath}
\setlength{\unitlength}{5mm}
\begin{picture}(8,4)(0,0.5)
\youngdg
\put(1,0){\makebox(1,1){$6$}}
\put(1,1){\makebox(1,1){$8$}}
\put(2,2){\makebox(1,1){$0$}}
\put(5.1,3){\makebox(1,1){$-5$}}
\put(7,1.5){\makebox(1,1){$\Rightarrow$}}
\end{picture}
\quad
\setlength{\unitlength}{5mm}
\begin{picture}(8,4)(0,0.5)
\youngdg
\put(1,0){\makebox(1,1){$7$}}
\put(1,1){\makebox(1,1){$9$}}
\put(2,2){\makebox(1,1){$2$}}
\put(5,3){\makebox(1,1){$0$}}
\put(7,1.5){\makebox(1,1){$\Rightarrow$}}
\end{picture}
\quad
\setlength{\unitlength}{5mm}
\begin{picture}(6,4)(0,0.5)
\youngdg
\put(1,0){\makebox(1,1){$8$}}
\put(1,1){\makebox(1,1){$10$}}
\put(2,2){\makebox(1,1){$4$}}
\put(5,3){\makebox(1,1){$5$}}
\end{picture}.
\end{displaymath}
The rows of the Young diagrams are regarded as
the {\em action} variables
and the numbers (riggings) associated with each row
the {\em angle} variables.
As one sees the former are invariant under the time
evolution while the latter increase by
the lengths of the associated rows.
In other words the time evolution is
linear in these variables.

In this paper
we present a new method to construct
a rigged configuration from a given state of
the box-ball system.
%Added on 25 May 
Here the procedure to construct a Young diagram
is the same as that in \cite{TTS,YYT} but
our method can extract additional data, the riggings.
Having the additional data
our method provides a bijection between the set of
automaton states and the set of rigged configurations.
To borrow a name from the theory of nonlinear evolution 
equations \cite{AC}
we call this bijection the 
{\em inverse scattering transform}
for the box-ball system, where we
regard the rigged configuration as a set of scattering data.
Using the technique of the inverse scattering transform
we give a proof of the assertion in \cite{KOTY}
that the rigged configuration
is indeed playing the role of a set of 
action-angle variables.
%%%%%%%%%%%%%%%%%%%%%%%%%%%%%%%%%%%%%%%%%%%%%%%%
\subsection{The box-ball system and the fermionic formulas}
We observe that a certain crystallization of
solvable lattice models can produce soliton cellular automata
\cite{HIK,HKT,HHIKTT}, while
there is a connection between solvable lattice
models and Bethe ansatz in a character
level \cite{HKOTY1,HKOTT}.
Now the inverse scattering transform for the box-ball system is
in some sense providing
a direct connection between soliton cellular
automata and Bethe ansatz.
The second purpose of this paper is to 
complete a study on this connection 
in the character level which was also proposed in \cite{KOTY}.

%%%%%%%%%%%%%%%%%%%%%%%%%%%%
Let us consider a particular case where
the character formula
comes from two different expressions for Kostka
polynomials \cite{Ma}.
One is from the Bethe ansatz \cite{KR} and yields
the fermionic character formula which is made of
sums of products of $q$-binomial coefficients.
The other
is from the solvable lattice models in
statistical mechanics \cite{NY} and yields 
the one-dimensional configuration sums
over paths (tensor products of crystals).
To be more specific we
consider the Kostka polynomial
$K_{\mu,\nu}(q)$ with $\mu=(L-s,s)$ and $\nu = (1^L)$
where $L$ and $s$ are two integers obeying
$L/2 \geq s \geq 0$.
Then the two expressions for the Kostka polynomial
lead to the identity \cite{HKOTY1}
\begin{align}
K_{\mu,\nu}(q) &=
\sum_{\lambda \vdash s} 
q^{\phi(\lambda)-L \sum_{i \geq 1} m_i(\lambda)+
\frac{L(L-1)}{2}}
\prod_{i \geq 1} { p_i(\lambda) + m_i(\lambda) 
\brack  m_i(\lambda) } \nonumber\\
&=\sum_{\mathbf{b} \in P_{L,s}}
q^{\sum_{j=1}^{L-1} (L-j) (1- \theta(b_j < b_{j+1}))},
\label{eq:may28_5}
\end{align}
where $\lambda \vdash s$ means that $\lambda$ is a
partition of $s$.
Here
$p_i(\lambda), m_i(\lambda)$ and $\phi(\lambda)$
will be defined in the main text
(equations (\ref{eq:may28_2}),(\ref{eq:may28_3})),
and $\theta(\mbox{true})=1, \theta(\mbox{false})=0$;
the set of highest weight paths is defined as
\begin{displaymath}
P_{L,s}=\left\{
\mathbf{b} = (b_1,\ldots,b_L) \Bigg\vert
b_i \in \{0,1\},
\sum_{j=1}^i b_j \leq \left\lfloor 
\frac{i}{2} \right\rfloor \,
\mbox{for any $i$}, 
\sum_{j=1}^L b_j = s
\right\}.
\end{displaymath}
There is a decomposition
$ P_{L,s} = \sqcup_{\lambda \vdash s} P_L(\lambda)$
such that if $\mathbf{b} \in P_L(\lambda)$ then
the relation $\sum_{j=1}^{L-1} \theta(b_j < b_{j+1}) =
\sum_{i \geq 1} m_i(\lambda)$ holds.
Therefore equation (\ref{eq:may28_5}) leads to
\begin{displaymath}
\sum_{\lambda \vdash s} 
\left( 
q^{\phi(\lambda)}
\prod_{i \geq 1} { p_i(\lambda) + m_i(\lambda) 
\brack  m_i(\lambda) } -
\sum_{\mathbf{b} \in P_{L}(\lambda)}
q^{\sum_{j=1}^{L-1} j \theta(b_j < b_{j+1})}
\right)=0.
\end{displaymath}
%%%%%%%%%%%%%%%%%%%%%%
Furthermore the decomposition makes each term of 
the sum $\sum_{\lambda \vdash s}$ vanish separately.
Thus there is actually a set of refined identities
behind (\ref{eq:may28_5}) with respect to $\lambda$.
%%%%%%%%%%%%%%%%%%%%%%%%%%%%%%%%%%%%
In the context of Bethe ansatz
this $\lambda$ has a clear meaning; it is a label
for the associated eigenstate of the Heisenberg magnet.
However in the context of solvable lattice models its meaning
is so far unclear.
By identifying the space of paths for a lattice
model (the six-vertex model in ferro-magnetic regime) with
the space of states of the box-ball system, we shall find
that the path associated with
$\lambda =(1^{m_1} 2^{m_2} \ldots)$ has 
$m_i$ ``solitons" of length $i$ for any $i$.
This interpretation is based on
an expression for
partition functions for the box-ball system
with specified soliton contents, which turns out to be
the above mentioned refined identities of the fermionic formula.
This expression for partition functions and its variant for
including non-highest weight paths were
proposed in \cite{KOTY}.
We derive these identities 
(Theorem \ref{prop:may26_3})
as a result of the inverse scattering transform
for the box-ball system.
%%%%%%%%%%%%%%%%%%%%%%%%%%%%%%%%%%%%%%%%%%%%%%
\subsection{Outline of the paper}
In Section~\ref{sec:2}
the box-ball system is introduced and its
updating rule is explained in terms of arcs.
In Section~\ref{sec:3} we define a mapping $\varXi$
which sends any state of the 
box-ball system to a rigged configuration (set of scattering data).
The inverse map of $\varXi$ (inverse scattering transform) is 
defined in Section~\ref{sec:4}.
The notion of wave tails and wave fronts of automaton states
is also introduced here.
In Section~\ref{sec:5} we give a proof that the time evolution
is linearized in the scattering data , and present an
inverse scattering method for this system.

In Section~\ref{sec:6} we derive formulas for
the right-most wave front
and the left-most wave tail of the automaton state
associated with a given rigged configuration.
In Section~\ref{sec:7} automaton states in a finite 
interval are studied and the notion of highest weight states
is introduced.
By taking summations over these states
we define partition functions for the box-ball system
in Section~\ref{sec:8} and establish the
conjectured fermionic formulas proposed in \cite{KOTY}.
In Section~\ref{sec:9} we prove that our inverse scattering
transform is equivalent to the $sl(2)$ case of the
bijection in \cite{KKR}.

In \ref{app:a} we give a proof and an explanation
of two formulas for
one-dimensional configuration sums expressed by $q$-binomial
coefficients.
%%%%%%%%%%%%%%%%%%%%%%%%%%%%%%%%%%%%%%%%%%%%%%%%%%
\section{The box-ball system}\label{sec:2}
We consider a one-dimensional array of 
infinite number of boxes that 
extends towards both directions.
As a position coordinate
we put successive integers to the {\em walls} between the
boxes rather than to the {\em boxes} themselves.

Any box is either an empty box or a filled box.
The latter means that there is a ball within the box.
We assume that there are at most finite number of balls
in the system.
The empty box is denoted by $\emp$ and the
filled box is by $\fil$.
Clearly there are four types of configurations of adjacent boxes,
$\empemp,\,\empfil,\,\filemp$ and $\filfil$.
We adopt the updating rule of the box-ball system in 
\cite{YYT}:
\begin{enumerate}
\item For every $\filemp$ connect its two boxes with an arc.
\item Ignore those
boxes connected with the arcs and
regard the other boxes as if they were successively adjoining.
\item Repeat steps 1 and 2 as many times as possible.
\item For every pair of connected boxes
interchange $\fil$ and $\emp$.
\end{enumerate}

\begin{example}
\label{ex:apr21_1}
We have
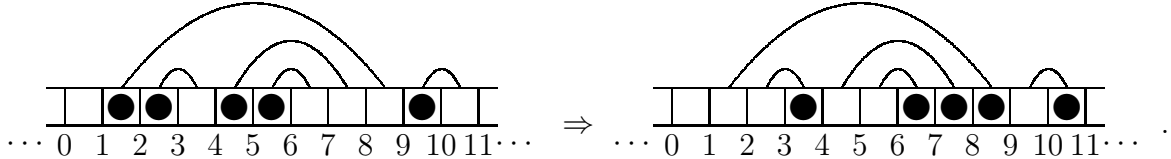
\begin{displaymath}
\setlength{\unitlength}{5mm}
\begin{picture}(14,3.3)(0,0.8)
\put(0.5,1){\line(1,0){12}}
\put(0.5,2){\line(1,0){12}}
\multiput(1,1)(1,0){12}{\line(0,1){1}}
\put(-0.5,0){\makebox(1,1){$\cdots$}}
\put(0.5,0){\makebox(1,1){$0$}}
\put(1.5,0){\makebox(1,1){$1$}}
\put(2.5,0){\makebox(1,1){$2$}}
\put(3.5,0){\makebox(1,1){$3$}}
\put(4.5,0){\makebox(1,1){$4$}}
\put(5.5,0){\makebox(1,1){$5$}}
\put(6.5,0){\makebox(1,1){$6$}}
\put(7.5,0){\makebox(1,1){$7$}}
\put(8.5,0){\makebox(1,1){$8$}}
\put(9.5,0){\makebox(1,1){$9$}}
\put(10.5,0){\makebox(1,1){$10$}}
\put(11.5,0){\makebox(1,1){$11$}}
\put(12.5,0){\makebox(1,1){$\cdots$}}
\qbezier[350](2.5,2)(6,6.5)(9.5,2)
\qbezier[80](3.5,2)(4,3)(4.5,2)
\qbezier[80](6.5,2)(7,3)(7.5,2)
\qbezier[80](10.5,2)(11,3)(11.5,2)
\qbezier[200](5.5,2)(7,4.5)(8.5,2)
%
% \put(1.5,1.5){\circle{0.7}}
\put(2.5,1.5){\circle*{0.7}}
\put(3.5,1.5){\circle*{0.7}}
% \put(4.5,1.5){\circle{0.7}}
\put(5.5,1.5){\circle*{0.7}}
\put(6.5,1.5){\circle*{0.7}}
% \put(7.5,1.5){\circle{0.7}}
% \put(8.5,1.5){\circle{0.7}}
% \put(9.5,1.5){\circle{0.7}}
\put(10.5,1.5){\circle*{0.7}}
% \put(11.5,1.5){\circle{0.7}}
\end{picture}
\Rightarrow 
\quad
\setlength{\unitlength}{5mm}
\begin{picture}(14,3.3)(0,0.8)
\put(0.5,1){\line(1,0){12}}
\put(0.5,2){\line(1,0){12}}
\multiput(1,1)(1,0){12}{\line(0,1){1}}
\put(-0.5,0){\makebox(1,1){$\cdots$}}
\put(0.5,0){\makebox(1,1){$0$}}
\put(1.5,0){\makebox(1,1){$1$}}
\put(2.5,0){\makebox(1,1){$2$}}
\put(3.5,0){\makebox(1,1){$3$}}
\put(4.5,0){\makebox(1,1){$4$}}
\put(5.5,0){\makebox(1,1){$5$}}
\put(6.5,0){\makebox(1,1){$6$}}
\put(7.5,0){\makebox(1,1){$7$}}
\put(8.5,0){\makebox(1,1){$8$}}
\put(9.5,0){\makebox(1,1){$9$}}
\put(10.5,0){\makebox(1,1){$10$}}
\put(11.5,0){\makebox(1,1){$11$}}
\put(12.5,0){\makebox(1,1){$\cdots$}}
\qbezier[350](2.5,2)(6,6.5)(9.5,2)
\qbezier[80](3.5,2)(4,3)(4.5,2)
\qbezier[80](6.5,2)(7,3)(7.5,2)
\qbezier[80](10.5,2)(11,3)(11.5,2)
\qbezier[200](5.5,2)(7,4.5)(8.5,2)
%
% \put(1.5,1.5){\circle{0.7}}
% \put(2.5,1.5){\circle*{0.7}}
% \put(3.5,1.5){\circle*{0.7}}
\put(4.5,1.5){\circle*{0.7}}
% \put(5.5,1.5){\circle*{0.7}}
% \put(6.5,1.5){\circle*{0.7}}
\put(7.5,1.5){\circle*{0.7}}
\put(8.5,1.5){\circle*{0.7}}
\put(9.5,1.5){\circle*{0.7}}
% \put(10.5,1.5){\circle*{0.7}}
\put(11.5,1.5){\circle*{0.7}}
\end{picture}.
\end{displaymath}
\end{example}
\begin{definition}[advanced/retarded arcs]
The arcs
that one will obtain by applying the items 1
(resp.~1 but $\filemp$ was replaced by $\empfil$), 
2, and 3 in the above procedure
to any automaton state
are called the {\em advanced arcs}
(resp.~{\em retarded arcs} ) of that state.
\end{definition}
By definition we have that
\begin{lemma}\label{lem:jun11_1}
Let $p$ and $p'$ be two automaton states.
Suppose the set of
advanced arcs of $p$ is equal to the 
set of retarded arcs of $p'$.
Then $p'$ is the state that one will obtain from $p$ by
updating the system once.
\end{lemma}
For later use we introduce the notion of
\begin{definition}[depth]
If an arc has no arc within it, its {\em depth} is set to
be one.
If an arc has arc(s) within it, its depth is given by
$1 + \max\{ \mbox{depths of inside arc(s)}\}$.
\end{definition}
For instance there are three depth 1, one depth 2,
and one depth 3 arcs in each figure of Example \ref{ex:apr21_1}.
%%%%%%%%%%%%%%%%%%%%%%%%%%%%%%%%%%%%%%%%%%%%%%%%%%%%%%%%%%%
\section{Scattering data and soliton contents}\label{sec:3}
Given a set of integers $\{\alpha_i\}_{1 \leq i \leq 2n}$
we introduce a matrix of the form
\begin{equation} \label{eq:apr23_1}
M = \left(
\begin{array}{llll}
\alpha_1 & \alpha_3 & \cdots & \alpha_{2n-1}  \\
\alpha_2 & \alpha_4 & \cdots & \alpha_{2n} 
\end{array}
\right).
\end{equation}
Let $\H_n$ (resp.~$\Hz_n$) be the set of all matrices 
of this form subject to the condition 
$\alpha_1 \leq \alpha_2 \leq \cdots \leq \alpha_{2n}$
(resp.~$\alpha_1 < \alpha_2 < \cdots < \alpha_{2n}$).
We define $\H := \sqcup_{n=0}^\infty \H_n$
(resp.~$\Hz := \sqcup_{n=0}^\infty \Hz_n$).
Here the set $\H_0 = \Hz_0$ consists of a formal two-row
matrix with no element.
We associate any matrix $M \in \Hz_n$ with
an automaton state that has
balls between the walls
$\alpha_{2k-1}$ and $\alpha_{2k}$ for $1 \leq k \leq n$.
We shall occasionally identity such a matrix $M$ with
the associated automaton state itself.
\begin{example}
\label{ex:apr19_2}
We have the identification
\begin{displaymath}
\setlength{\unitlength}{5mm}
\begin{picture}(14,1.7)(0,1)
\put(0.5,1){\line(1,0){12}}
\put(0.5,2){\line(1,0){12}}
\multiput(1,1)(1,0){12}{\line(0,1){1}}
\put(-0.5,0){\makebox(1,1){$\cdots$}}
\put(0.5,0){\makebox(1,1){$0$}}
\put(1.5,0){\makebox(1,1){$1$}}
\put(2.5,0){\makebox(1,1){$2$}}
\put(3.5,0){\makebox(1,1){$3$}}
\put(4.5,0){\makebox(1,1){$4$}}
\put(5.5,0){\makebox(1,1){$5$}}
\put(6.5,0){\makebox(1,1){$6$}}
\put(7.5,0){\makebox(1,1){$7$}}
\put(8.5,0){\makebox(1,1){$8$}}
\put(9.5,0){\makebox(1,1){$9$}}
\put(10.5,0){\makebox(1,1){$10$}}
\put(11.5,0){\makebox(1,1){$11$}}
\put(12.5,0){\makebox(1,1){$\cdots$}}
%
% \put(1.5,1.5){\circle{0.7}}
\put(2.5,1.5){\circle*{0.7}}
\put(3.5,1.5){\circle*{0.7}}
% \put(4.5,1.5){\circle{0.7}}
\put(5.5,1.5){\circle*{0.7}}
\put(6.5,1.5){\circle*{0.7}}
% \put(7.5,1.5){\circle{0.7}}
% \put(8.5,1.5){\circle{0.7}}
% \put(9.5,1.5){\circle{0.7}}
\put(10.5,1.5){\circle*{0.7}}
% \put(11.5,1.5){\circle{0.7}}
\end{picture}
\leftrightarrow 
\quad
M = \left(
\begin{array}{ccc}
1 & 4 & 9 \\
3 & 6 & 10
\end{array}
\right).
\end{displaymath}
\end{example}

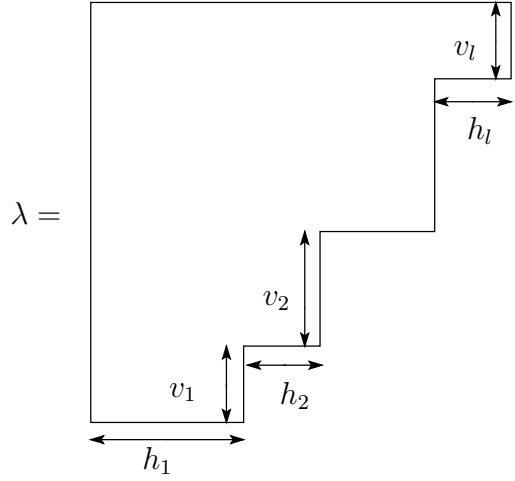
\begin{figure}
%WinTpicVersion3.08
\unitlength 0.1in
% \begin{picture}( 26.2000, 23.0000)( 11.8000,-31.0000)
\begin{picture}( 26.2000, 23.0000)( 0.8000,-31.0000)
% POLYGON 2 0 3 0
% 13 1600 800 1600 3000 2400 3000 2400 2600 2800 2600 2800 2000 3400 2000 3400 1200 3800 1200 3800 800 3800 800 3800 800 1600 800
% 
\special{pn 8}%
\special{pa 1600 800}%
\special{pa 1600 3000}%
\special{pa 2400 3000}%
\special{pa 2400 2600}%
\special{pa 2800 2600}%
\special{pa 2800 2000}%
\special{pa 3400 2000}%
\special{pa 3400 1200}%
\special{pa 3800 1200}%
\special{pa 3800 800}%
\special{pa 3800 800}%
\special{pa 3800 800}%
\special{pa 1600 800}%
\special{fp}%
% VECTOR 2 0 3 0
% 2 2000 3090 2400 3090
% 
\special{pn 8}%
\special{pa 2000 3090}%
\special{pa 2400 3090}%
\special{fp}%
\special{sh 1}%
\special{pa 2400 3090}%
\special{pa 2334 3070}%
\special{pa 2348 3090}%
\special{pa 2334 3110}%
\special{pa 2400 3090}%
\special{fp}%
% STR 2 0 3 0
% 3 1870 3170 1870 3270 2 0
% $h_1$
\put(18.7000,-32.7000){\makebox(0,0)[lb]{$h_1$}}%
% VECTOR 2 0 3 0
% 2 2000 3090 1600 3090
% 
\special{pn 8}%
\special{pa 2000 3090}%
\special{pa 1600 3090}%
\special{fp}%
\special{sh 1}%
\special{pa 1600 3090}%
\special{pa 1668 3110}%
\special{pa 1654 3090}%
\special{pa 1668 3070}%
\special{pa 1600 3090}%
\special{fp}%
% VECTOR 2 0 3 0
% 2 2610 2700 2810 2700
% 
\special{pn 8}%
\special{pa 2610 2700}%
\special{pa 2810 2700}%
\special{fp}%
\special{sh 1}%
\special{pa 2810 2700}%
\special{pa 2744 2680}%
\special{pa 2758 2700}%
\special{pa 2744 2720}%
\special{pa 2810 2700}%
\special{fp}%
% VECTOR 2 0 3 0
% 2 2620 2700 2420 2700
% 
\special{pn 8}%
\special{pa 2620 2700}%
\special{pa 2420 2700}%
\special{fp}%
\special{sh 1}%
\special{pa 2420 2700}%
\special{pa 2488 2720}%
\special{pa 2474 2700}%
\special{pa 2488 2680}%
\special{pa 2420 2700}%
\special{fp}%
% VECTOR 2 0 3 0
% 4 3600 1320 3800 1320 3600 1320 3400 1320
% 
\special{pn 8}%
\special{pa 3600 1320}%
\special{pa 3800 1320}%
\special{fp}%
\special{sh 1}%
\special{pa 3800 1320}%
\special{pa 3734 1300}%
\special{pa 3748 1320}%
\special{pa 3734 1340}%
\special{pa 3800 1320}%
\special{fp}%
\special{pa 3600 1320}%
\special{pa 3400 1320}%
\special{fp}%
\special{sh 1}%
\special{pa 3400 1320}%
\special{pa 3468 1340}%
\special{pa 3454 1320}%
\special{pa 3468 1300}%
\special{pa 3400 1320}%
\special{fp}%
% VECTOR 2 0 3 0
% 4 2310 2800 2310 3000 2310 2820 2310 2600
% 
\special{pn 8}%
\special{pa 2310 2800}%
\special{pa 2310 3000}%
\special{fp}%
\special{sh 1}%
\special{pa 2310 3000}%
\special{pa 2330 2934}%
\special{pa 2310 2948}%
\special{pa 2290 2934}%
\special{pa 2310 3000}%
\special{fp}%
\special{pa 2310 2820}%
\special{pa 2310 2600}%
\special{fp}%
\special{sh 1}%
\special{pa 2310 2600}%
\special{pa 2290 2668}%
\special{pa 2310 2654}%
\special{pa 2330 2668}%
\special{pa 2310 2600}%
\special{fp}%
% VECTOR 2 0 3 0
% 8 2720 2400 2720 2600 2720 2400 2720 2000 3720 1000 3720 1200 3720 1000 3720 800
% 
\special{pn 8}%
\special{pa 2720 2400}%
\special{pa 2720 2600}%
\special{fp}%
\special{sh 1}%
\special{pa 2720 2600}%
\special{pa 2740 2534}%
\special{pa 2720 2548}%
\special{pa 2700 2534}%
\special{pa 2720 2600}%
\special{fp}%
\special{pa 2720 2400}%
\special{pa 2720 2000}%
\special{fp}%
\special{sh 1}%
\special{pa 2720 2000}%
\special{pa 2700 2068}%
\special{pa 2720 2054}%
\special{pa 2740 2068}%
\special{pa 2720 2000}%
\special{fp}%
\special{pa 3720 1000}%
\special{pa 3720 1200}%
\special{fp}%
\special{sh 1}%
\special{pa 3720 1200}%
\special{pa 3740 1134}%
\special{pa 3720 1148}%
\special{pa 3700 1134}%
\special{pa 3720 1200}%
\special{fp}%
\special{pa 3720 1000}%
\special{pa 3720 800}%
\special{fp}%
\special{sh 1}%
\special{pa 3720 800}%
\special{pa 3700 868}%
\special{pa 3720 854}%
\special{pa 3740 868}%
\special{pa 3720 800}%
\special{fp}%
% STR 2 0 3 0
% 3 2590 2830 2590 2930 2 0
% $h_2$
\put(25.9000,-29.3000){\makebox(0,0)[lb]{$h_2$}}%
% STR 2 0 3 0
% 3 3570 1440 3570 1540 2 0
% $h_l$
\put(35.7000,-15.4000){\makebox(0,0)[lb]{$h_l$}}%
% STR 2 0 3 0
% 3 2010 2790 2010 2890 2 0
% $v_1$
\put(20.1000,-28.9000){\makebox(0,0)[lb]{$v_1$}}%
% STR 2 0 3 0
% 3 2500 2310 2500 2410 2 0
% $v_2$
\put(25.0000,-24.1000){\makebox(0,0)[lb]{$v_2$}}%
% STR 2 0 3 0
% 3 3500 1000 3500 1100 2 0
% $v_l$
\put(35.0000,-11.0000){\makebox(0,0)[lb]{$v_l$}}%
% STR 2 0 3 0
% 3 1180 1860 1180 1960 2 0
% $\lambda = $
\put(11.8000,-19.6000){\makebox(0,0)[lb]{$\lambda = $}}%
\end{picture}%
\caption{The Young diagram.}
\label{fig:1}
\end{figure}

We show a procedure that associates a rigged configuration,
a Young diagram with a set of integers, for any
$M \in \Hz$.
Suppose $M \in \Hz_n$ and
let $n_1 = n$ and $M_1 = M$.
Given any non-negative integer $i$ and
a matrix of the form
\begin{equation} \label{eq:apr19_1}
M_i = \left(
\begin{array}{llll}
\alpha_1^i & \alpha_3^i & \cdots & \alpha_{2n_i-1}^i  \\
\alpha_2^i & \alpha_4^i & \cdots & \alpha_{2n_i}^i 
\end{array}
\right) \in \Hz_{n_{i}}
\end{equation}
we define a set of integers
$h_i,v_i,a_1^i,\ldots,a_{v_i}^i$ 
(the upper indices are not exponents) and
a matrix $M_{i+1} \in \Hz_{n_{i+1}}$.
Here $n_{i+1} = n_i - v_i$.
We set $h_i = \min_{1 \leq k \leq 2n_i-1}
(\alpha_{k+1}^i - \alpha_{k}^i)$ and 
$\beta_k^i = \alpha_k^i - k h_i$.
{}From the set of integers $\{ \beta_k^i \}_{1 \leq k \leq
2 n_i}$ we remove pairs of identical integers repeatedly
until there is no such pair.
Let $v_i = \#\mbox{(removed pairs)}$.
We arrange the removed integers 
(after their multiplicity was divided by two)
in increasing order
and call them $a_1^i,\ldots,a_{v_i}^i$; we call
the remaining integers 
$\alpha_1^{i+1},\ldots,\alpha_{2 n_{i+1}}^{i+1}$
in increasing order.
Then let $M_{i+1}$ be a matrix of the form
(\ref{eq:apr19_1}) but
with all the $i$'s were replaced by $i+1$'s.
By repeating this we shall obtain
a matrix $M_{l+1}$ which has no element
after a finite number $(l)$ of steps.
Then let 
$\lambda$ be the Young diagram in Figure \ref{fig:1}
and 
% $a=\{\{a_j^i\}_{1 \leq j \leq v_i}\}_{1 \leq i \leq l}$
$a=\sqcup_{1 \leq i \leq l}\{a_j^i\}_{1 \leq j \leq v_i}$
the set of integers obtained through this procedure.

We denote $\R$ (resp.~$\P$)
the set of all rigged configurations
(resp.~Young diagrams).
Let $\varXi: \Hz \rightarrow \R$ 
(resp.~$\xi: \Hz \rightarrow \P$)
be the mapping that sends
any matrix $M \in \Hz$ to a
rigged configuration $\varXi(M)=(\lambda,a) \in \R$
(resp.~a Young diagram $\xi(M) = \lambda \in \P$)
by the above procedure.
For a reason that will be clear afterwards we call them
\begin{definition}[scattering data, soliton content]
\label{def:may26_2}
For any $M \in \Hz$ (or the automaton state associated with
$M$) the rigged configuration $\varXi(M)$ is called its 
{\em scattering data}, and
the Young diagram $\xi(M)$ is called its
{\em soliton content}.
\end{definition}
The mapping $\varXi$ becomes a bijection
between $\Hz$ and $\R$.
Its inverse map will be discussed in the next section.

\begin{example}\label{ex:may28_1}
We consider the matrix $M$ in Example \ref{ex:apr19_2}.
Then we have 
$h_1 = \min \{3-1,4-3,6-4,9-6,10-9 \} =1$.
Since
\begin{displaymath}
\left(
\begin{array}{ccc}
1 & 4 & 9 \\
3 & 6 & 10
\end{array}
\right) -
\left(
\begin{array}{rrr}
1 & 3 & 5 \\
2 & 4 & 6
\end{array}
\right) \times 1 =
\left(
\begin{array}{rrr}
0 & \textit{1} & \textit{4} \\
\textit{1} & 2 & \textit{4}
\end{array}
\right),
\end{displaymath}
we obtain $v_1=2,a^1_1=1,a^1_2=4$, and
$M_2 = \left(
\begin{array}{r}
0  \\
2 
\end{array}
\right)$.
Clearly $h_2 = 2$ and since
\begin{displaymath}
\left(
\begin{array}{r}
0  \\
2 
\end{array}
\right) -
\left(
\begin{array}{r}
1  \\
2 
\end{array}
\right) \times 2 =
\left(
\begin{array}{r}
\textit{-2}  \\
\textit{-2} 
\end{array}
\right),
\end{displaymath}
we obtain $v_2=1$ and $a^2_1=-2$.
Therefore we have
\begin{displaymath}
\left(
\begin{array}{rrr}
1 & 4 & 9 \\
3 & 6 & 10
\end{array}
\right)  \quad
\stackrel{\displaystyle \varXi}{\mapsto} 
\quad
\setlength{\unitlength}{5mm}
\begin{picture}(4,2.5)(0,1)
\put(0,0){\line(1,0){1}}
\put(0,1){\line(1,0){1}}
\put(0,2){\line(1,0){3}}
\put(0,3){\line(1,0){3}}
\put(0,0){\line(0,1){3}}
\put(1,0){\line(0,1){3}}
\put(2,2){\line(0,1){1}}
\put(3,2){\line(0,1){1}}
\put(1,0){\makebox(1,1){$1$}}
\put(1,1){\makebox(1,1){$4$}}
\put(3.1,2){\makebox(1,1){$-2$}}
\end{picture}.
\end{displaymath}
\end{example}
\vspace{1cm}
We usually depict a rigged configuration by
putting the riggings along the associated
vertical edges of the Young diagram
(See Example \ref{ex:may28_1}).
This convention matches to the following
step by step description of
the mapping $\varXi$.
Suppose we are on a square lattice made of vertices
connected by unit length bonds, 
and have a matrix $M \in \Hz$.
We start at an arbitrary vertex of the lattice.
Every step we proceed rightward or upward by one unit length 
and replace our matrix $M$ by the following rule.
First suppose $M \notin \Hz_0$.
\begin{enumerate}
\item If $M \in \Hz$ we proceed rightward and replace $M$ by 
$M - { 1 3 \cdots \choose 2 4 \cdots }$ that is in $\H$.
We place a horizontal segment of unit length on the way.
\item 
If $M \in \H$ but $M \notin \Hz$ there are pairs of same 
integers in $M$.
Find a pair of smallest same integers, one of which is in
the first row and the other in the second row.
We proceed upward and remove the pair from $M$.
We place a vertical segment of unit length on the way and
put the removed integer on its right-hand side.
\end{enumerate}
We repeat this procedure until our matrix $M$ has no more entry.
Then we stop there.
By connecting the segments we obtain a path of lower-left
to upper-right direction.
By attaching an upper left corner to it (to
complete a Young diagram) we obtain the rigged configuration.
%%%%%%%%%%%%%%%%%%%%%%%%%%%%%%%%%%%%%%%%%%%%%%%%%%%
\section{Inverse scattering transform} \label{sec:4}
In this section we study the inverse of the mapping $\varXi$.
For any integer $x$ 
we define a map $\Omega_x:\H \rightarrow \H$
by the following rule.
Given $M \in \H_n$ we label its
elements as (\ref{eq:apr23_1}).
Let $i$ be the largest integer obeying the condition
$\alpha_i \leq x$; if $\alpha_1 > x$ we let $i=0$.
For an odd $i(=2l-1)$ we set
\begin{displaymath}
\Omega_x(M) := \left(
\begin{array}{llllll}
\alpha_1 & \cdots & \alpha_{2l-1} & x& \cdots & \alpha_{2n-1}  \\
\alpha_2 & \cdots & x & \alpha_{2l} & \cdots & \alpha_{2n} 
\end{array}
\right),
\end{displaymath}
and for an even $i(=2l)$
\begin{displaymath}
\Omega_x(M) := \left(
\begin{array}{lllllll}
\alpha_1 & \cdots & \alpha_{2l-1} & x & 
\alpha_{2l+1} & \cdots & \alpha_{2n-1}  \\
\alpha_2 & \cdots & \alpha_{2l} & x & 
\alpha_{2l+2} & \cdots & \alpha_{2n} 
\end{array}
\right).
\end{displaymath}
We extend our definition
to any set of integers $X$.
Given $M \in \H$ and $X=\{x_1,\ldots,x_p \}$
we set
\begin{displaymath}
\Omega_X (M) := \Omega_{x_1} \circ \cdots \circ
\Omega_{x_p} (M).
\end{displaymath}
In addition we define the following maps 
$\Phi_i:\H \rightarrow \Hz \quad (i=0,1)$.
Given $M \in \H_n$ we set
\begin{align*}
\Phi_0(M) &= M + \left(
\begin{array}{llll}
0 & 2 & \cdots & 2n-2  \\
1 & 3 & \cdots & 2n-1
\end{array}
\right), 
\Phi_1(M) = M + \left(
\begin{array}{llll}
1 & 3 & \cdots & 2n-1  \\
2 & 4 & \cdots & 2n
\end{array}
\right).
\end{align*}
For later analyses it is important to notice that
the automaton state associated with
$\Phi_{1} \circ \Omega_X(M)$ 
(resp.~$\Phi_{0} \circ \Omega_X(M)$)
% for $X=\{x_1,\ldots,x_p\}$ and $M \in \H_n$
is made by the following algorithm.
We begin with the automaton state associated with $M$.
\begin{enumerate}
\item At the wall positions
$x_1,\ldots,x_p$ and 
$\alpha_1,\alpha_3,\ldots,\alpha_{2n-1}$
(resp.~$\alpha_2,\alpha_4,\ldots,\alpha_{2n}$), put marks, say $\vee$.
Here their multiplicity should be taken into account.
\item Split the array of boxes 
at the left-most wall with the mark(s).
If it has $p(\geq 1)$ mark(s), translate every
box and ball on
the right-hand side of the wall
rightwards by the width of $2p$ boxes.
Delete the mark(s) at the wall.
\item Repeat item 2 as many times as possible.
\item Fill $\empfil$s 
(resp.~ $\filemp$s) into the gaps
between the arrays of boxes made by the above
procedures.
\end{enumerate}
\begin{remark}\label{rem:apr30_3}
This algorithm also applies to $\Phi_0$ and $\Phi_1$
themselves, because
we can regard them as $\Phi_0 \circ \Omega_X$ 
and $\Phi_1 \circ \Omega_X$ with $X = \emptyset$.
\end{remark}
\begin{example}\label{ex:may14_4}
Let $M$ be the matrix in Example \ref{ex:apr19_2} and
let $X=\{1,5,8\}$.
Then we have
\begin{align*}
\Phi_{1} \circ \Omega_X(M) &=
\left(
\begin{array}{cccccc}
\textit{1} & 1 & 4 & \textit{5} & \textit{8} & 9  \\
\textit{1} & 3 & \textit{5} & 6 & \textit{8} & 10
\end{array}
\right)+
\left(
\begin{array}{cccccc}
1 & 3 & 5 & 7 & 9 & 11  \\
2 & 4 & 6 & 8 & 10 & 12
\end{array}
\right) \\
&=
\left(
\begin{array}{cccccc}
2 & 4 & 9 & 12 & 17 & 20  \\
3 & 7 & 11 & 14 & 18 & 22
\end{array}
\right),\\
\intertext{and}
\Phi_{0} \circ \Omega_X(M) &=
\left(
\begin{array}{cccccc}
1 & 3 & 8 & 11 & 16 & 19  \\
2 & 6 & 10 & 13 & 17 & 21
\end{array}
\right).
\end{align*}
The algorithm to make the automaton state associated with
$\Phi_{1} \circ \Omega_X(M)$
is depicted as
\begin{align*}
&
\setlength{\unitlength}{5mm}
\begin{picture}(14,2)(0,0.5)
\put(0.5,1){\line(1,0){12}}
\put(0.5,2){\line(1,0){12}}
\multiput(1,1)(1,0){12}{\line(0,1){1}}
\put(-0.5,0){\makebox(1,1){$\cdots$}}
\put(0.5,0){\makebox(1,1){$0$}}
\put(1.5,0){\makebox(1,1){$1$}}
\put(1.5,2){\makebox(1,1){$\vee$}}
\put(1.5,2.5){\makebox(1,1){$\vee$}}
\put(2.5,0){\makebox(1,1){$2$}}
\put(3.5,0){\makebox(1,1){$3$}}
\put(4.5,0){\makebox(1,1){$4$}}
\put(4.5,2){\makebox(1,1){$\vee$}}
\put(5.5,0){\makebox(1,1){$5$}}
\put(5.5,2){\makebox(1,1){$\vee$}}
\put(6.5,0){\makebox(1,1){$6$}}
\put(7.5,0){\makebox(1,1){$7$}}
\put(8.5,0){\makebox(1,1){$8$}}
\put(8.5,2){\makebox(1,1){$\vee$}}
\put(9.5,0){\makebox(1,1){$9$}}
\put(9.5,2){\makebox(1,1){$\vee$}}
\put(10.5,0){\makebox(1,1){$10$}}
\put(11.5,0){\makebox(1,1){$11$}}
\put(12.5,0){\makebox(1,1){$\cdots$}}
\put(2.5,1.5){\circle*{0.7}}
\put(3.5,1.5){\circle*{0.7}}
\put(5.5,1.5){\circle*{0.7}}
\put(6.5,1.5){\circle*{0.7}}
\put(10.5,1.5){\circle*{0.7}}
\end{picture}
\\
&\qquad \qquad \qquad \qquad \downarrow
\\
&
\setlength{\unitlength}{5mm}
\begin{picture}(22,2)(0,0.5)
\put(0.5,1){\line(1,0){1.5}}
\put(6,1){\line(1,0){3}}
\put(11,1){\line(1,0){1}}
\put(14,1){\line(1,0){3}}
\put(19,1){\line(1,0){1}}
\put(22,1){\line(1,0){1.5}}
\put(0.5,2){\line(1,0){1.5}}
\put(6,2){\line(1,0){3}}
\put(11,2){\line(1,0){1}}
\put(14,2){\line(1,0){3}}
\put(19,2){\line(1,0){1}}
\put(22,2){\line(1,0){1.5}}
\multiput(1,1)(1,0){2}{\line(0,1){1}}
\multiput(6,1)(1,0){4}{\line(0,1){1}}
\multiput(11,1)(1,0){2}{\line(0,1){1}}
\multiput(14,1)(1,0){4}{\line(0,1){1}}
\multiput(19,1)(1,0){2}{\line(0,1){1}}
\multiput(22,1)(1,0){2}{\line(0,1){1}}
\put(-0.5,0){\makebox(1,1){$\cdots$}}
\put(0.5,0){\makebox(1,1){$0$}}
\put(1.5,0){\makebox(1,1){$1$}}
\put(2.5,0){\makebox(1,1){$2$}}
\put(3.5,0){\makebox(1,1){$3$}}
\put(4.5,0){\makebox(1,1){$4$}}
\put(5.5,0){\makebox(1,1){$5$}}
\put(6.5,0){\makebox(1,1){$6$}}
\put(7.5,0){\makebox(1,1){$7$}}
\put(8.5,0){\makebox(1,1){$8$}}
\put(9.5,0){\makebox(1,1){$9$}}
\put(10.5,0){\makebox(1,1){$10$}}
\put(11.5,0){\makebox(1,1){$11$}}
\put(12.5,0){\makebox(1,1){$12$}}
\put(13.5,0){\makebox(1,1){$13$}}
\put(14.5,0){\makebox(1,1){$14$}}
\put(15.5,0){\makebox(1,1){$15$}}
\put(16.5,0){\makebox(1,1){$16$}}
\put(17.5,0){\makebox(1,1){$17$}}
\put(18.5,0){\makebox(1,1){$18$}}
\put(19.5,0){\makebox(1,1){$19$}}
\put(20.5,0){\makebox(1,1){$20$}}
\put(21.5,0){\makebox(1,1){$21$}}
\put(22.5,0){\makebox(1,1){$22$}}
\put(23.5,0){\makebox(1,1){$\cdots$}}
\put(6.5,1.5){\circle*{0.7}}
\put(7.5,1.5){\circle*{0.7}}
\put(11.5,1.5){\circle*{0.7}}
\put(14.5,1.5){\circle*{0.7}}
\put(22.5,1.5){\circle*{0.7}}
\end{picture}
\\
&\qquad \qquad \qquad \qquad \downarrow
\\
&
\setlength{\unitlength}{5mm}
\begin{picture}(22,2)(0,0.5)
\put(0.5,1){\line(1,0){23}}
\put(0.5,2){\line(1,0){23}}
\multiput(1,1)(1,0){23}{\line(0,1){1}}
\put(-0.5,0){\makebox(1,1){$\cdots$}}
\put(0.5,0){\makebox(1,1){$0$}}
\put(1.5,0){\makebox(1,1){$1$}}
\put(2.5,0){\makebox(1,1){$2$}}
\put(3.5,0){\makebox(1,1){$3$}}
\put(4.5,0){\makebox(1,1){$4$}}
\put(5.5,0){\makebox(1,1){$5$}}
\put(6.5,0){\makebox(1,1){$6$}}
\put(7.5,0){\makebox(1,1){$7$}}
\put(8.5,0){\makebox(1,1){$8$}}
\put(9.5,0){\makebox(1,1){$9$}}
\put(10.5,0){\makebox(1,1){$10$}}
\put(11.5,0){\makebox(1,1){$11$}}
\put(12.5,0){\makebox(1,1){$12$}}
\put(13.5,0){\makebox(1,1){$13$}}
\put(14.5,0){\makebox(1,1){$14$}}
\put(15.5,0){\makebox(1,1){$15$}}
\put(16.5,0){\makebox(1,1){$16$}}
\put(17.5,0){\makebox(1,1){$17$}}
\put(18.5,0){\makebox(1,1){$18$}}
\put(19.5,0){\makebox(1,1){$19$}}
\put(20.5,0){\makebox(1,1){$20$}}
\put(21.5,0){\makebox(1,1){$21$}}
\put(22.5,0){\makebox(1,1){$22$}}
\put(23.5,0){\makebox(1.5,1){$\cdots$ .}}
\put(3.5,1.5){\circle*{0.7}}
\put(5.5,1.5){\circle*{0.7}}
\put(6.5,1.5){\circle*{0.7}}
\put(7.5,1.5){\circle*{0.7}}
\put(10.5,1.5){\circle*{0.7}}
\put(11.5,1.5){\circle*{0.7}}
\put(13.5,1.5){\circle*{0.7}}
\put(14.5,1.5){\circle*{0.7}}
\put(18.5,1.5){\circle*{0.7}}
\put(21.5,1.5){\circle*{0.7}}
\put(22.5,1.5){\circle*{0.7}}
\end{picture}
\end{align*}
%%%%%%%%%%%%%%%%%%%%%%%%%%%%
In contrast that for $\Phi_{0} \circ \Omega_X(M)$
is depicted as
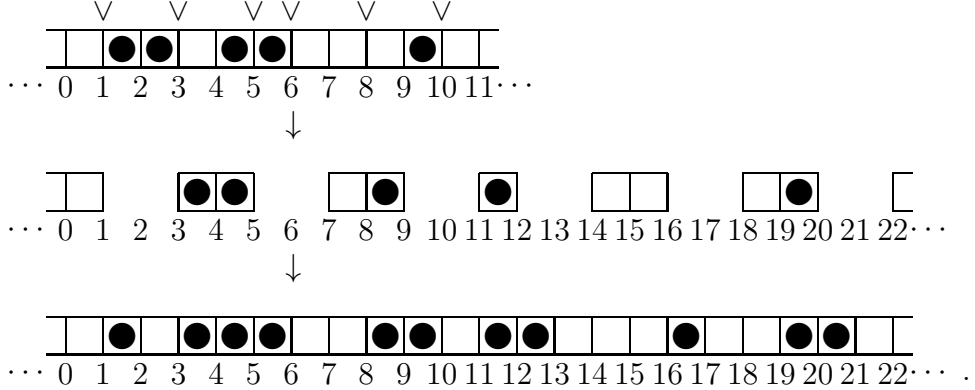
\begin{align*}
&
\setlength{\unitlength}{5mm}
\begin{picture}(14,2)(0,0.5)
\put(0.5,1){\line(1,0){12}}
\put(0.5,2){\line(1,0){12}}
\multiput(1,1)(1,0){12}{\line(0,1){1}}
\put(-0.5,0){\makebox(1,1){$\cdots$}}
\put(0.5,0){\makebox(1,1){$0$}}
\put(1.5,0){\makebox(1,1){$1$}}
\put(1.5,2){\makebox(1,1){$\vee$}}
\put(2.5,0){\makebox(1,1){$2$}}
\put(3.5,0){\makebox(1,1){$3$}}
\put(3.5,2){\makebox(1,1){$\vee$}}
\put(4.5,0){\makebox(1,1){$4$}}
\put(5.5,0){\makebox(1,1){$5$}}
\put(5.5,2){\makebox(1,1){$\vee$}}
\put(6.5,0){\makebox(1,1){$6$}}
\put(6.5,2){\makebox(1,1){$\vee$}}
\put(7.5,0){\makebox(1,1){$7$}}
\put(8.5,0){\makebox(1,1){$8$}}
\put(8.5,2){\makebox(1,1){$\vee$}}
\put(9.5,0){\makebox(1,1){$9$}}
\put(10.5,0){\makebox(1,1){$10$}}
\put(10.5,2){\makebox(1,1){$\vee$}}
\put(11.5,0){\makebox(1,1){$11$}}
\put(12.5,0){\makebox(1,1){$\cdots$}}
\put(2.5,1.5){\circle*{0.7}}
\put(3.5,1.5){\circle*{0.7}}
\put(5.5,1.5){\circle*{0.7}}
\put(6.5,1.5){\circle*{0.7}}
\put(10.5,1.5){\circle*{0.7}}
\end{picture}
\\
&\qquad \qquad \qquad \qquad \downarrow
\\
&
\setlength{\unitlength}{5mm}
\begin{picture}(22,2)(0,0.5)
\put(0.5,1){\line(1,0){1.5}}
\put(4,1){\line(1,0){2}}
\put(8,1){\line(1,0){2}}
\put(12,1){\line(1,0){1}}
\put(15,1){\line(1,0){2}}
\put(19,1){\line(1,0){2}}
\put(23,1){\line(1,0){0.5}}
\put(0.5,2){\line(1,0){1.5}}
\put(4,2){\line(1,0){2}}
\put(8,2){\line(1,0){2}}
\put(12,2){\line(1,0){1}}
\put(15,2){\line(1,0){2}}
\put(19,2){\line(1,0){2}}
\put(23,2){\line(1,0){0.5}}
\multiput(1,1)(1,0){2}{\line(0,1){1}}
\multiput(4,1)(1,0){3}{\line(0,1){1}}
\multiput(8,1)(1,0){3}{\line(0,1){1}}
\multiput(12,1)(1,0){2}{\line(0,1){1}}
\multiput(15,1)(1,0){3}{\line(0,1){1}}
\multiput(19,1)(1,0){3}{\line(0,1){1}}
\put(23,1){\line(0,1){1}}
\put(-0.5,0){\makebox(1,1){$\cdots$}}
\put(0.5,0){\makebox(1,1){$0$}}
\put(1.5,0){\makebox(1,1){$1$}}
\put(2.5,0){\makebox(1,1){$2$}}
\put(3.5,0){\makebox(1,1){$3$}}
\put(4.5,0){\makebox(1,1){$4$}}
\put(5.5,0){\makebox(1,1){$5$}}
\put(6.5,0){\makebox(1,1){$6$}}
\put(7.5,0){\makebox(1,1){$7$}}
\put(8.5,0){\makebox(1,1){$8$}}
\put(9.5,0){\makebox(1,1){$9$}}
\put(10.5,0){\makebox(1,1){$10$}}
\put(11.5,0){\makebox(1,1){$11$}}
\put(12.5,0){\makebox(1,1){$12$}}
\put(13.5,0){\makebox(1,1){$13$}}
\put(14.5,0){\makebox(1,1){$14$}}
\put(15.5,0){\makebox(1,1){$15$}}
\put(16.5,0){\makebox(1,1){$16$}}
\put(17.5,0){\makebox(1,1){$17$}}
\put(18.5,0){\makebox(1,1){$18$}}
\put(19.5,0){\makebox(1,1){$19$}}
\put(20.5,0){\makebox(1,1){$20$}}
\put(21.5,0){\makebox(1,1){$21$}}
\put(22.5,0){\makebox(1,1){$22$}}
\put(23.5,0){\makebox(1,1){$\cdots$}}
\put(4.5,1.5){\circle*{0.7}}
\put(5.5,1.5){\circle*{0.7}}
\put(9.5,1.5){\circle*{0.7}}
\put(12.5,1.5){\circle*{0.7}}
\put(20.5,1.5){\circle*{0.7}}
\end{picture}
\\
&\qquad \qquad \qquad \qquad \downarrow
\\
&
\setlength{\unitlength}{5mm}
\begin{picture}(22,2)(0,0.5)
\put(0.5,1){\line(1,0){23}}
\put(0.5,2){\line(1,0){23}}
\multiput(1,1)(1,0){23}{\line(0,1){1}}
\put(-0.5,0){\makebox(1,1){$\cdots$}}
\put(0.5,0){\makebox(1,1){$0$}}
\put(1.5,0){\makebox(1,1){$1$}}
\put(2.5,0){\makebox(1,1){$2$}}
\put(3.5,0){\makebox(1,1){$3$}}
\put(4.5,0){\makebox(1,1){$4$}}
\put(5.5,0){\makebox(1,1){$5$}}
\put(6.5,0){\makebox(1,1){$6$}}
\put(7.5,0){\makebox(1,1){$7$}}
\put(8.5,0){\makebox(1,1){$8$}}
\put(9.5,0){\makebox(1,1){$9$}}
\put(10.5,0){\makebox(1,1){$10$}}
\put(11.5,0){\makebox(1,1){$11$}}
\put(12.5,0){\makebox(1,1){$12$}}
\put(13.5,0){\makebox(1,1){$13$}}
\put(14.5,0){\makebox(1,1){$14$}}
\put(15.5,0){\makebox(1,1){$15$}}
\put(16.5,0){\makebox(1,1){$16$}}
\put(17.5,0){\makebox(1,1){$17$}}
\put(18.5,0){\makebox(1,1){$18$}}
\put(19.5,0){\makebox(1,1){$19$}}
\put(20.5,0){\makebox(1,1){$20$}}
\put(21.5,0){\makebox(1,1){$21$}}
\put(22.5,0){\makebox(1,1){$22$}}
\put(23.5,0){\makebox(1.5,1){$\cdots$ .}}
\put(2.5,1.5){\circle*{0.7}}
\put(4.5,1.5){\circle*{0.7}}
\put(5.5,1.5){\circle*{0.7}}
\put(6.5,1.5){\circle*{0.7}}
\put(9.5,1.5){\circle*{0.7}}
\put(10.5,1.5){\circle*{0.7}}
\put(12.5,1.5){\circle*{0.7}}
\put(13.5,1.5){\circle*{0.7}}
\put(17.5,1.5){\circle*{0.7}}
\put(20.5,1.5){\circle*{0.7}}
\put(21.5,1.5){\circle*{0.7}}
\end{picture}
\end{align*}
\end{example}
\vspace{1cm}
\begin{definition}[wave tails/fronts]
In the state of the box-ball system associated with
$M \in \Hz$ the wall positions specified by the
first row of $M$ are called {\em wave tails}.
The wall positions specified by the
second row of $M$ are called {\em wave fronts}.
\end{definition}
\begin{remark}\label{rem:apr30_2}
For any set of integers
$X$ the mapping $\Phi_{1} \circ \Omega_X$ 
(resp.~$\Phi_{0} \circ \Omega_X$)
embeds the box pair $\empfil$ (resp.~$\filemp$)
at every wave tail (resp.~wave front) of $M$, and
at every
wall position specified by $X$.
This implies that the set of 
retarded arcs of $\Phi_{1} \circ \Omega_X(M)$ 
(resp.~advanced arcs of $\Phi_{0} \circ \Omega_X(M)$) 
consists of ---
1) depth $\geq 2$ arcs made out of
the retarded arcs (resp.~advanced arcs) of $M$,
being stretched but their topology unchanged, and ---
2) depth $1$ arcs introduced
along with the newly embedded box pairs.
See Figure \ref{fig:2} which corresponds to the latter
case in Example \ref{ex:may14_4}.
\end{remark}
\begin{figure}[htbp]
\begin{align*}
&
\setlength{\unitlength}{5mm}
\begin{picture}(14,4)(0,0)
\put(0.5,1){\line(1,0){12}}
\put(0.5,2){\line(1,0){12}}
\multiput(1,1)(1,0){12}{\line(0,1){1}}
\put(-0.5,0){\makebox(1,1){$\cdots$}}
\put(0.5,0){\makebox(1,1){$0$}}
\put(1.5,0){\makebox(1,1){$1$}}
\put(2.5,0){\makebox(1,1){$2$}}
\put(3.5,0){\makebox(1,1){$3$}}
\put(4.5,0){\makebox(1,1){$4$}}
\put(5.5,0){\makebox(1,1){$5$}}
\put(6.5,0){\makebox(1,1){$6$}}
\put(7.5,0){\makebox(1,1){$7$}}
\put(8.5,0){\makebox(1,1){$8$}}
\put(9.5,0){\makebox(1,1){$9$}}
\put(10.5,0){\makebox(1,1){$10$}}
\put(11.5,0){\makebox(1,1){$11$}}
\put(12.5,0){\makebox(1,1){$\cdots$}}
\put(2.5,1.5){\circle*{0.7}}
\put(3.5,1.5){\circle*{0.7}}
\put(5.5,1.5){\circle*{0.7}}
\put(6.5,1.5){\circle*{0.7}}
\put(10.5,1.5){\circle*{0.7}}
\qbezier[350](2.5,2)(6,6.5)(9.5,2)
\qbezier[80](3.5,2)(4,3)(4.5,2)
\qbezier[80](6.5,2)(7,3)(7.5,2)
\qbezier[80](10.5,2)(11,3)(11.5,2)
\qbezier[200](5.5,2)(7,4.5)(8.5,2)
\end{picture}
\\
&
\\
&
\setlength{\unitlength}{5mm}
\begin{picture}(22,3)(0,0)
\put(0.5,1){\line(1,0){1.5}}
\put(4,1){\line(1,0){2}}
\put(8,1){\line(1,0){2}}
\put(12,1){\line(1,0){1}}
\put(15,1){\line(1,0){2}}
\put(19,1){\line(1,0){2}}
\put(23,1){\line(1,0){0.5}}
\put(0.5,2){\line(1,0){1.5}}
\put(4,2){\line(1,0){2}}
\put(8,2){\line(1,0){2}}
\put(12,2){\line(1,0){1}}
\put(15,2){\line(1,0){2}}
\put(19,2){\line(1,0){2}}
\put(23,2){\line(1,0){0.5}}
\multiput(1,1)(1,0){2}{\line(0,1){1}}
\multiput(4,1)(1,0){3}{\line(0,1){1}}
\multiput(8,1)(1,0){3}{\line(0,1){1}}
\multiput(12,1)(1,0){2}{\line(0,1){1}}
\multiput(15,1)(1,0){3}{\line(0,1){1}}
\multiput(19,1)(1,0){3}{\line(0,1){1}}
\put(23,1){\line(0,1){1}}
\put(-0.5,0){\makebox(1,1){$\cdots$}}
\put(0.5,0){\makebox(1,1){$0$}}
\put(1.5,0){\makebox(1,1){$1$}}
\put(2.5,0){\makebox(1,1){$2$}}
\put(3.5,0){\makebox(1,1){$3$}}
\put(4.5,0){\makebox(1,1){$4$}}
\put(5.5,0){\makebox(1,1){$5$}}
\put(6.5,0){\makebox(1,1){$6$}}
\put(7.5,0){\makebox(1,1){$7$}}
\put(8.5,0){\makebox(1,1){$8$}}
\put(9.5,0){\makebox(1,1){$9$}}
\put(10.5,0){\makebox(1,1){$10$}}
\put(11.5,0){\makebox(1,1){$11$}}
\put(12.5,0){\makebox(1,1){$12$}}
\put(13.5,0){\makebox(1,1){$13$}}
\put(14.5,0){\makebox(1,1){$14$}}
\put(15.5,0){\makebox(1,1){$15$}}
\put(16.5,0){\makebox(1,1){$16$}}
\put(17.5,0){\makebox(1,1){$17$}}
\put(18.5,0){\makebox(1,1){$18$}}
\put(19.5,0){\makebox(1,1){$19$}}
\put(20.5,0){\makebox(1,1){$20$}}
\put(21.5,0){\makebox(1,1){$21$}}
\put(22.5,0){\makebox(1,1){$22$}}
\put(23.5,0){\makebox(1,1){$\cdots$}}
\put(4.5,1.5){\circle*{0.7}}
\put(5.5,1.5){\circle*{0.7}}
\put(9.5,1.5){\circle*{0.7}}
\put(12.5,1.5){\circle*{0.7}}
\put(20.5,1.5){\circle*{0.7}}
\qbezier[600](4.5,2)(10,7.5)(19.5,2)
\qbezier[200](5.5,2)(7,4.5)(8.5,2)
\qbezier[350](9.5,2)(14,5.5)(16.5,2)
\qbezier[200](12.5,2)(14,4.5)(15.5,2)
\qbezier[200](20.5,2)(22,4.5)(23.5,2)
\end{picture}
\\
&
\\
&
\setlength{\unitlength}{5mm}
\begin{picture}(22,3)(0,0)
\put(0.5,1){\line(1,0){23}}
\put(0.5,2){\line(1,0){23}}
\multiput(1,1)(1,0){23}{\line(0,1){1}}
\put(-0.5,0){\makebox(1,1){$\cdots$}}
\put(0.5,0){\makebox(1,1){$0$}}
\put(1.5,0){\makebox(1,1){$1$}}
\put(2.5,0){\makebox(1,1){$2$}}
\put(3.5,0){\makebox(1,1){$3$}}
\put(4.5,0){\makebox(1,1){$4$}}
\put(5.5,0){\makebox(1,1){$5$}}
\put(6.5,0){\makebox(1,1){$6$}}
\put(7.5,0){\makebox(1,1){$7$}}
\put(8.5,0){\makebox(1,1){$8$}}
\put(9.5,0){\makebox(1,1){$9$}}
\put(10.5,0){\makebox(1,1){$10$}}
\put(11.5,0){\makebox(1,1){$11$}}
\put(12.5,0){\makebox(1,1){$12$}}
\put(13.5,0){\makebox(1,1){$13$}}
\put(14.5,0){\makebox(1,1){$14$}}
\put(15.5,0){\makebox(1,1){$15$}}
\put(16.5,0){\makebox(1,1){$16$}}
\put(17.5,0){\makebox(1,1){$17$}}
\put(18.5,0){\makebox(1,1){$18$}}
\put(19.5,0){\makebox(1,1){$19$}}
\put(20.5,0){\makebox(1,1){$20$}}
\put(21.5,0){\makebox(1,1){$21$}}
\put(22.5,0){\makebox(1,1){$22$}}
\put(23.5,0){\makebox(1,1){$\cdots$}}
\put(2.5,1.5){\circle*{0.7}}
\put(4.5,1.5){\circle*{0.7}}
\put(5.5,1.5){\circle*{0.7}}
\put(6.5,1.5){\circle*{0.7}}
\put(9.5,1.5){\circle*{0.7}}
\put(10.5,1.5){\circle*{0.7}}
\put(12.5,1.5){\circle*{0.7}}
\put(13.5,1.5){\circle*{0.7}}
\put(17.5,1.5){\circle*{0.7}}
\put(20.5,1.5){\circle*{0.7}}
\put(21.5,1.5){\circle*{0.7}}
\qbezier[600](4.5,2)(10,7.5)(19.5,2)
\qbezier[200](5.5,2)(7,4.5)(8.5,2)
\qbezier[350](9.5,2)(14,5.5)(16.5,2)
\qbezier[200](12.5,2)(14,4.5)(15.5,2)
\qbezier[200](20.5,2)(22,4.5)(23.5,2)
\qbezier[80](2.5,2)(3,3)(3.5,2)
\qbezier[80](6.5,2)(7,3)(7.5,2)
\qbezier[80](10.5,2)(11,3)(11.5,2)
\qbezier[80](13.5,2)(14,3)(14.5,2)
\qbezier[80](17.5,2)(18,3)(18.5,2)
\qbezier[80](21.5,2)(22,3)(22.5,2)
\end{picture}
\end{align*}
\caption{The embedding of box pairs and 
the advanced arcs.}
\label{fig:2}
\end{figure}
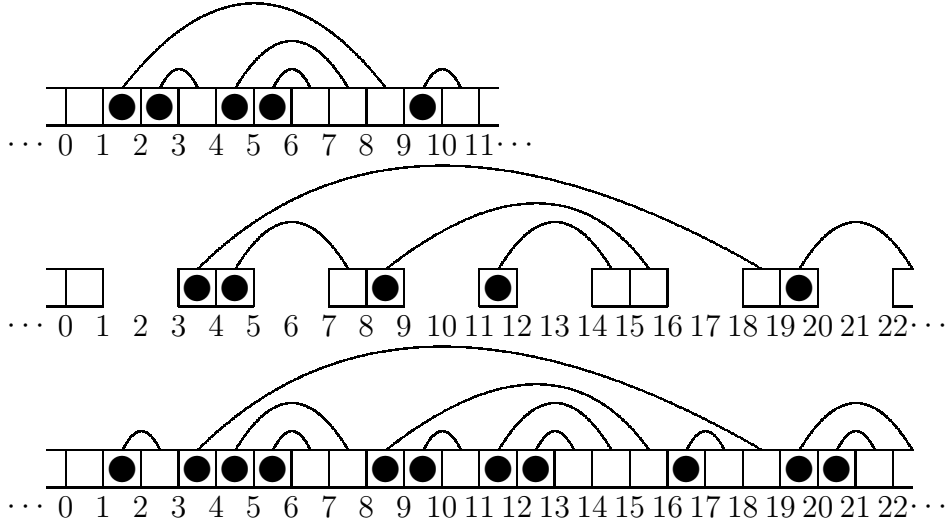
%
%%%%%%%%%%%%%%%%%%%%%%%%%%%%%%%%%%%%%%%%%%%%%
Let $(\lambda,a)$ be a rigged configuration where
$\lambda$ is the Young diagram in Figure \ref{fig:1}
and $a=\sqcup_{1 \leq i \leq l}\{a_j^i\}_{1 \leq j \leq v_i}$
a set of integers.
For any $i$ we denote $\{a_j^i\}_{1 \leq j \leq v_i}$ by $a^i$
(Again, the upper indices are not exponents).
Let $ L_{l+1}(\lambda,a) = M_{l+1}(\lambda,a) \in \Hz_0$ 
be the two-row matrix that has
no element.
For any $i \, (1 \leq i \leq l)$ 
we define the matrices 
$L_i(\lambda,a), M_i(\lambda,a) \in \Hz_{n_i}$ 
(where $n_i = \sum_{k=i}^l v_k$) recursively as
\begin{align*}
L_i(\lambda,a) &:=
(\Phi_0)^{h_i} \circ \Omega_{a^i} (L_{i+1}(\lambda,a)),\\
M_i(\lambda,a) &:=
(\Phi_1)^{h_i} \circ \Omega_{a^i} (M_{i+1}(\lambda,a)).
\end{align*}
We write
$L(\lambda,a) := L_1(\lambda,a)$ and
$M(\lambda,a) := M_1(\lambda,a)$.
Then we can regard $L$ and $M$ as mappings from $\R$ to $\Hz$.
It is easy to see that 
the map $M$ is the inverse of the map
$\varXi$ in Section~\ref{sec:3}.
\begin{proposition}\label{prop:may14_2}
$\varXi (M(\lambda,a)) = (\lambda,a)$.
\end{proposition}
We call this map
\begin{definition}[inverse scattering transform]
For any $(\lambda,a) \in \R$ the mapping (or its image)
$(\lambda,a) \mapsto M(\lambda,a) \in \Hz$ is called its
{\em inverse scattering transform}.
\end{definition}
The inverse scattering transform and
the mapping $\varXi$ (which may be called direct
scattering transform)
form a bijection between $\R$ and $\Hz$.
We will also call this bijection itself 
the inverse scattering transform for the 
box-ball system.
%
%%%%%%%%%%%%%%%%%%%%%%%%%%%%%%%%%%%%%%%%%%%%%%%%
\section{A proof of the linearized time evolution of
the scattering data}\label{sec:5}
The generic appearance of the automaton states
in the {\em real space} changes in not a simple way
under the time evolution.
According to the time evolution
the set of associated scattering data also evolves.
In this section we show that the time evolution of the
scattering data is rather simple;
it turns out to be linear.

Recall the two-row matrices $M_i(\lambda,a)$ and
$L_i(\lambda,a)$ in Section \ref{sec:4}.
For any $i$ we have
\begin{lemma}\label{lem:apr26_2}
The first row of $M_i(\lambda,a)$ is
equal to the second row of $L_i(\lambda,a)$. 
\end{lemma}
\begin{proof}
We give a proof by descending induction on $i$.
For $i=l+1$ the claim holds since the matrices 
have no entry.
Let $(\beta_1,\ldots,\beta_p)$ be the first row of
$M_{i+1}(\lambda,a)$ and suppose it is equal to
the second row of $L_{i+1}(\lambda,a)$.
Then the first row of $\Omega_{a^i}(M_{i+1}(\lambda,a))$
is equal to the second row of
$\Omega_{a^i}(L_{i+1}(\lambda,a))$ since
both of them are made out of
$\{ a^i_1,\ldots,a^i_{v_i},\beta_1,\ldots,\beta_p \}$
by rearranging them in increasing order.
The claim of the lemma follows immediately because of
the definitions of the mappings $\Phi_0$ and $\Phi_1$.
\end{proof}

For any $i$ we have
\begin{lemma}\label{lem:may27_8}
The set of advanced arcs of $L_i(\lambda,a)$ is equal to the
set of retarded arcs of $M_i(\lambda,a)$.
\end{lemma}
\begin{proof}
We give a proof by descending induction on $i$.
For $i=l+1$ the claim holds since the matrices 
have no entry.
Suppose that the claim is true if $i$ was replaced by $i+1$.
First we will show the following assertion:
\begin{itemize}
\item The set of advanced arcs of
$\Phi_0 \circ \Omega_{a^i}(L_{i+1}(\lambda,a))$ is
equal to the set of retarded arcs of
$\Phi_1 \circ \Omega_{a^i}(M_{i+1}(\lambda,a))$.
\end{itemize}
Here the
$\Phi_0 \circ \Omega_{a^i}$
embeds $\filemp$s to the wave fronts of 
$L_{i+1}(\lambda,a)$ 
and the wall positions specified by the set $a^i$;
the $\Phi_1 \circ \Omega_{a^i}$
embeds $\empfil$s to the wave tails of 
$M_{i+1}(\lambda,a)$ 
and the wall positions specified by the same $a^i$.
But the wave fronts of
$L_{i+1}(\lambda,a)$ are the wave tails of
$M_{i+1}(\lambda,a)$ by Lemma \ref{lem:apr26_2}.
Therefore every embedding point coincides
exactly between both cases.
This together with the assumption of the induction
concludes that the above assertion is true
(See Remark \ref{rem:apr30_2}).
By repeating this argument we can obtain
the claim of the lemma for $i$ 
(See Remark \ref{rem:apr30_3}).
The proof is completed.
\end{proof}
By Lemma~\ref{lem:jun11_1} it leads to
\begin{corollary}\label{cor:apr30_4}
If the state of the box-ball system associated with
$L(\lambda,a)$ is updated once, it is
the state of the system associated with
$M(\lambda,a)$.
\end{corollary}
We give an example for Lemmas \ref{lem:apr26_2} and
\ref{lem:may27_8}.
%
%%%%%%%%%%%%%%%%%%%%%%%%%%%%%%%%%%%%%%%%%%%%%%
\begin{example}\label{ex:may27_6}
Recall the second rigged configuration
of the  example in Section~\ref{sec:1}.
\begin{displaymath}
\setlength{\unitlength}{5mm}
\begin{picture}(8,4)(0,0.5)
\youngdg
\put(1,0){\makebox(1,1){$7$}}
\put(1,1){\makebox(1,1){$9$}}
\put(2,2){\makebox(1,1){$2$}}
\put(5,3){\makebox(1,1){$0$}}
\put(-3,1.5){\makebox(2,1){$(\lambda,a)=$}}
\end{picture}
\end{displaymath}
The $\lambda$ has the shape of
$h_1=h_2=1,h_3=3$ and
$v_1=2,v_2=v_3=1$.
The set of riggings is
$a = a^1 \sqcup a^2 \sqcup a^3$ with
$a^1 = \{ 7,9 \}, a^2 = \{ 2 \}, a^3 = \{ 0 \}$.
Thus we have 
(by denoting $L_i(\lambda,a)$ by $L_i$)
\begin{align*}
&\Phi_0 \circ \Omega_{a^3}(L_4) = {0 \choose 1},
\quad
\Phi_0^2 \circ \Omega_{a^3}(L_4) = {0 \choose 2},
\quad
L_3 = \Phi_0^3 \circ \Omega_{a^3}(L_4) = 
{0 \choose 3},\\
&L_2 = \Phi_0 \circ \Omega_{a^2}(L_3) =
\left(
\begin{array}{ll}
0 & \textit{2} \\
\textit{2} & 3 
\end{array}
\right) + 
\left(
\begin{array}{ll}
0 & 2 \\
1 & 3 
\end{array}
\right) =
\left(
\begin{array}{ll}
0 & 4 \\
3 & 6 
\end{array}
\right), 
\\
&L_1 = \Phi_0 \circ \Omega_{a^1}(L_2) =
\left(
\begin{array}{llll}
0 & 4 & \textit{7} & \textit{9}  \\
3 & 6 & \textit{7} & \textit{9} 
\end{array}
\right) +
\left(
\begin{array}{llll}
0 & 2 & 4 & 6 \\
1 & 3 & 5 & 7 
\end{array}
\right) =
\left(
\begin{array}{llll}
0 & 6 & 11 & 15 \\
4 & 9 & 12 & 16 
\end{array}
\right).
\end{align*}
Their advanced arcs are given as follows.
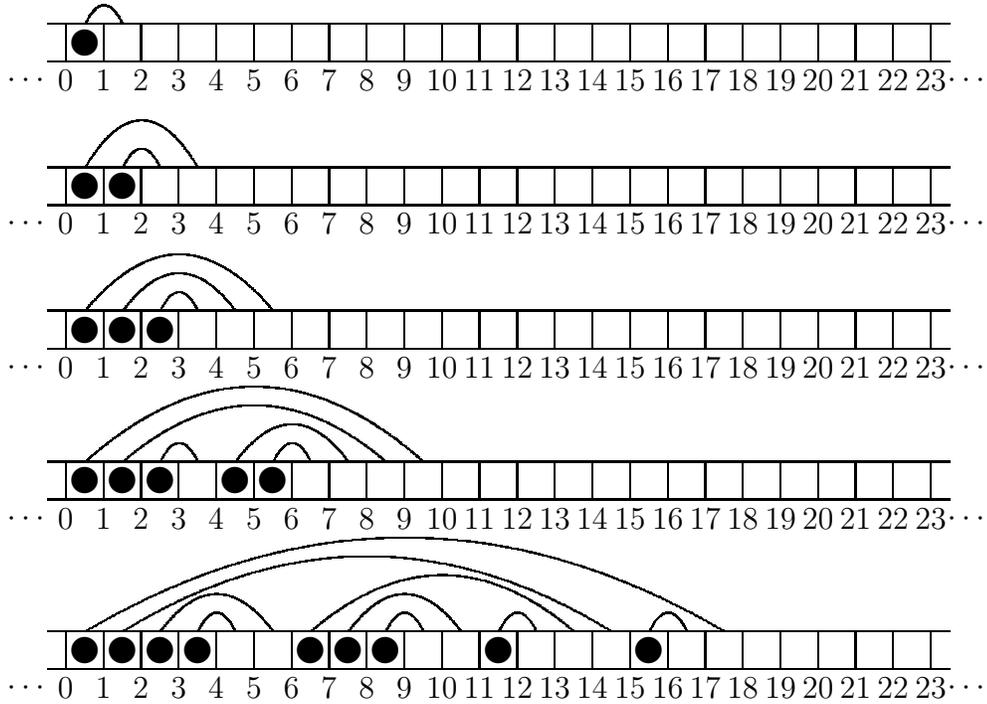
\begin{align*}
&
\setlength{\unitlength}{5mm}
\begin{picture}(23,2)(0,0.5)
\boxarray
\put(1.5,1.5){\circle*{0.7}}
\qbezier[80](1.5,2)(2,3)(2.5,2)
\end{picture}
\\
&
\\
&
\setlength{\unitlength}{5mm}
\begin{picture}(23,2)(0,0.5)
\boxarray
\put(1.5,1.5){\circle*{0.7}}
\put(2.5,1.5){\circle*{0.7}}
\qbezier[80](2.5,2)(3,3)(3.5,2)
\qbezier[200](1.5,2)(3,4.5)(4.5,2)
\end{picture}
\\
&
\\
&
\setlength{\unitlength}{5mm}
\begin{picture}(23,2)(0,0.5)
\boxarray
\put(1.5,1.5){\circle*{0.7}}
\put(2.5,1.5){\circle*{0.7}}
\put(3.5,1.5){\circle*{0.7}}
\qbezier[80](3.5,2)(4,3)(4.5,2)
\qbezier[200](2.5,2)(4,4)(5.5,2)
\qbezier[300](1.5,2)(4,5)(6.5,2)
\end{picture}
\\
&
\\
&
\setlength{\unitlength}{5mm}
\begin{picture}(23,2.2)(0,0.5)
\boxarray
\put(1.5,1.5){\circle*{0.7}}
\put(2.5,1.5){\circle*{0.7}}
\put(3.5,1.5){\circle*{0.7}}
\put(5.5,1.5){\circle*{0.7}}
\put(6.5,1.5){\circle*{0.7}}
\qbezier[80](3.5,2)(4,3)(4.5,2)
\qbezier[80](6.5,2)(7,3)(7.5,2)
\qbezier[200](5.5,2)(7,4)(8.5,2)
\qbezier[300](2.5,2)(6,5)(9.5,2)
\qbezier[400](1.5,2)(6,6)(10.5,2)
\end{picture}
\\
&
\\
&
\setlength{\unitlength}{5mm}
\begin{picture}(23,2.7)(0,0.5)
\boxarray
\put(1.5,1.5){\circle*{0.7}}
\put(2.5,1.5){\circle*{0.7}}
\put(3.5,1.5){\circle*{0.7}}
\put(4.5,1.5){\circle*{0.7}}
\put(7.5,1.5){\circle*{0.7}}
\put(8.5,1.5){\circle*{0.7}}
\put(9.5,1.5){\circle*{0.7}}
\put(12.5,1.5){\circle*{0.7}}
\put(16.5,1.5){\circle*{0.7}}
\qbezier[80](4.5,2)(5,3)(5.5,2)
\qbezier[80](9.5,2)(10,3)(10.5,2)
\qbezier[80](12.5,2)(13,3)(13.5,2)
\qbezier[80](16.5,2)(17,3)(17.5,2)
\qbezier[200](3.5,2)(5,4)(6.5,2)
\qbezier[200](8.5,2)(10,4)(11.5,2)
\qbezier[300](7.5,2)(11,5)(14.5,2)
\qbezier[400](2.5,2)(9,6)(15.5,2)
\qbezier[500](1.5,2)(10,7)(18.5,2)
\end{picture}
\end{align*}
%%%%%%%%%%%%%%%%%%%%%%%%%%%%%%%%%%%%%%%%%%%%%%%%
On the other hand we have
(by denoting $M_i(\lambda,a)$ by $M_i$)
\begin{align*}
&\Phi_1 \circ \Omega_{a^3}(M_4) = {1 \choose 2},
\quad
\Phi_1^2 \circ \Omega_{a^3}(M_4) = {2 \choose 4},
\quad
M_3 = \Phi_1^3 \circ \Omega_{a^3}(M_4) = 
{3 \choose 6},\\
&M_2 = \Phi_1 \circ \Omega_{a^2}(M_3) =
\left(
\begin{array}{ll}
\textit{2} & 3 \\
\textit{2} & 6 
\end{array}
\right) + 
\left(
\begin{array}{ll}
1 & 3 \\
2 & 4 
\end{array}
\right) =
\left(
\begin{array}{cc}
3 & 6 \\
4 & 10 
\end{array}
\right), 
\\
&M_1 = \Phi_1 \circ \Omega_{a^1}(M_2) =
\left(
\begin{array}{cccc}
3 & 6 & \textit{7} & \textit{9}  \\
4 & \textit{7} & \textit{9} & 10
\end{array}
\right) +
\left(
\begin{array}{cccc}
1 & 3 & 5 & 7 \\
2 & 4 & 6 & 8
\end{array}
\right) =
\left(
\begin{array}{cccc}
4 & 9 & 12 & 16 \\
6 & 11 & 15 & 18 
\end{array}
\right).
\end{align*}
%%%%%%%%%%%%%%%%%%%%%%%%%%%%%%%%%%%%%%%%%%%%%%%%%%
Their retarded arcs are given as follows.
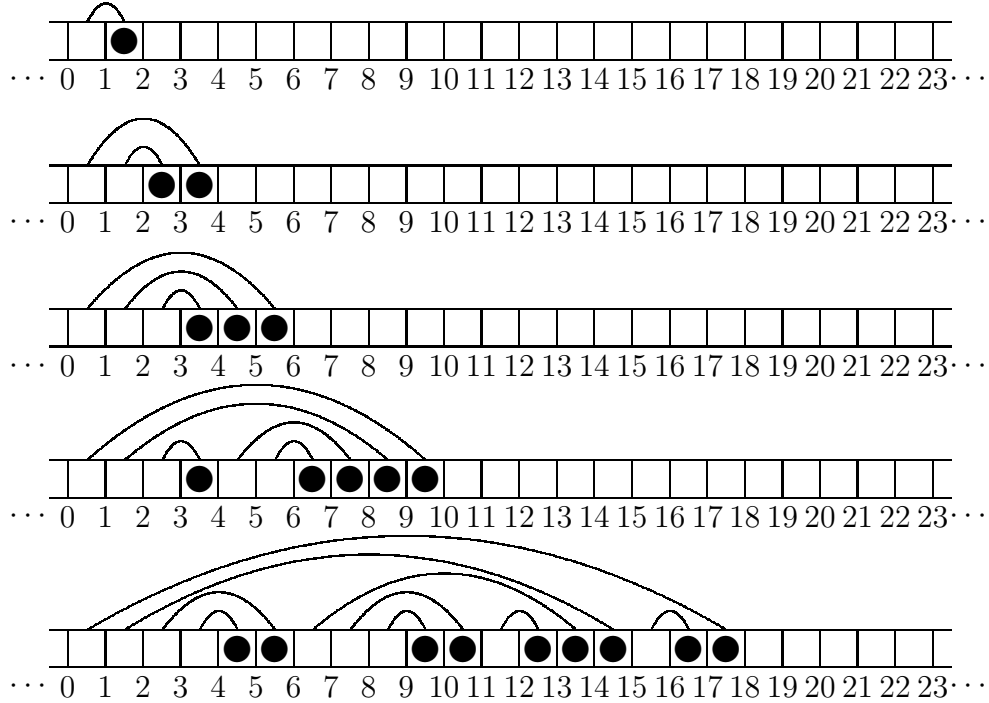
\begin{align*}
&
\setlength{\unitlength}{5mm}
\begin{picture}(23,2)(0,0.5)
\boxarray
\put(2.5,1.5){\circle*{0.7}}
\qbezier[80](1.5,2)(2,3)(2.5,2)
\end{picture}
\\
&
\\
&
\setlength{\unitlength}{5mm}
\begin{picture}(23,2)(0,0.5)
\boxarray
\put(3.5,1.5){\circle*{0.7}}
\put(4.5,1.5){\circle*{0.7}}
\qbezier[80](2.5,2)(3,3)(3.5,2)
\qbezier[200](1.5,2)(3,4.5)(4.5,2)
\end{picture}
\\
&
\\
&
\setlength{\unitlength}{5mm}
\begin{picture}(23,2)(0,0.5)
\boxarray
\put(4.5,1.5){\circle*{0.7}}
\put(5.5,1.5){\circle*{0.7}}
\put(6.5,1.5){\circle*{0.7}}
\qbezier[80](3.5,2)(4,3)(4.5,2)
\qbezier[200](2.5,2)(4,4)(5.5,2)
\qbezier[300](1.5,2)(4,5)(6.5,2)
\end{picture}
\\
&
\\
&
\setlength{\unitlength}{5mm}
\begin{picture}(23,2.2)(0,0.5)
\boxarray
\put(4.5,1.5){\circle*{0.7}}
\put(7.5,1.5){\circle*{0.7}}
\put(8.5,1.5){\circle*{0.7}}
\put(9.5,1.5){\circle*{0.7}}
\put(10.5,1.5){\circle*{0.7}}
\qbezier[80](3.5,2)(4,3)(4.5,2)
\qbezier[80](6.5,2)(7,3)(7.5,2)
\qbezier[200](5.5,2)(7,4)(8.5,2)
\qbezier[300](2.5,2)(6,5)(9.5,2)
\qbezier[400](1.5,2)(6,6)(10.5,2)
\end{picture}
\\
&
\\
&
\setlength{\unitlength}{5mm}
\begin{picture}(23,2.7)(0,0.5)
\boxarray
\put(5.5,1.5){\circle*{0.7}}
\put(6.5,1.5){\circle*{0.7}}
\put(10.5,1.5){\circle*{0.7}}
\put(11.5,1.5){\circle*{0.7}}
\put(13.5,1.5){\circle*{0.7}}
\put(14.5,1.5){\circle*{0.7}}
\put(15.5,1.5){\circle*{0.7}}
\put(17.5,1.5){\circle*{0.7}}
\put(18.5,1.5){\circle*{0.7}}
\qbezier[80](4.5,2)(5,3)(5.5,2)
\qbezier[80](9.5,2)(10,3)(10.5,2)
\qbezier[80](12.5,2)(13,3)(13.5,2)
\qbezier[80](16.5,2)(17,3)(17.5,2)
\qbezier[200](3.5,2)(5,4)(6.5,2)
\qbezier[200](8.5,2)(10,4)(11.5,2)
\qbezier[300](7.5,2)(11,5)(14.5,2)
\qbezier[400](2.5,2)(9,6)(15.5,2)
\qbezier[500](1.5,2)(10,7)(18.5,2)
\end{picture}
\end{align*}
%%%%%%%%%%%%%%%%%%%%%%%%%%%%%%%%%%%
\end{example}
\vspace{1cm}
%%%%%%%%%%%%%%%%%%%%%%%%%%%%%%%%%%%%%%%%%%%%
Let $\underline{a}$ be the rigging made out of $a$
by replacing $a^i_j$ by $\underline{a}^i_j:=a^i_j - 
\sum_{k=1}^i h_k$.
Let $\underline{a}^i$ denote the set
$\{  \underline{a}^i_j\}_{1 \leq j \leq v_i}$.
We set $T_1 :=\Phi_0^{-1} \circ \Phi_1$.
By the same symbol we write as
$T_1(X) = \{x_1+1,\ldots,x_p+1\}$
for any set of integers $X=\{x_1,\ldots,x_p\}$.
For instance we have ${a}^i=
T_1^{h_1+\cdots+h_{i}}(\underline{a}^i)$.
It is easy to see that 
\begin{lemma}\label{lem:may7_1}
For any $M \in \H$ and any set of integers $X$
the relation
$\Omega_{T_1(X)}(T_1(M)) = T_1 \circ \Omega_X(M)$
holds.
\end{lemma}
Then for any $i$ we have
\begin{lemma}\label{lem:may27_9}
The following relation holds:
$L_i(\lambda,{a}) = T_1^{h_1+\cdots+h_{i-1}} 
(M_i(\lambda,\underline{a}))$.
\end{lemma}
\begin{proof}
We give a proof by descending induction on $i$.
For $i=l+1$ the claim holds because both sides are
equal to a matrix with no entry.
Suppose 
$L_{i+1}(\lambda,{a}) = T_1^{h_1+\cdots+h_{i}} 
(M_{i+1}(\lambda,\underline{a}))$.
By repeated use of
Lemma \ref{lem:may7_1} we have
$\Omega_{{a}^i}(L_{i+1}(\lambda,{a})) = 
T_1^{h_1+\cdots+h_{i}} \circ
\Omega_{\underline{a}^i}(M_{i+1}(\lambda,\underline{a}))$.
Then
\begin{align*}
L_i(\lambda,{a}) &= \Phi_0^{h_i} \circ
\Omega_{{a}^i}(L_{i+1}(\lambda,{a})) \\
&= \Phi_0^{h_i} \circ
(\Phi_0^{-1} \circ \Phi_1)^{h_1+\cdots+h_{i}} \circ
\Omega_{\underline{a}^i}(M_{i+1}(\lambda,\underline{a})) \\
&= (\Phi_0^{-1} \circ \Phi_1)^{h_1+\cdots+h_{i-1}} \circ
\Phi_1^{h_i} \circ \Omega_{\underline{a}^i}
(M_{i+1}(\lambda,\underline{a})) \\
&=  T_1^{h_1+\cdots+h_{i-1}} (M_i(\lambda,\underline{a})).
\end{align*}
\end{proof}
By setting $i=1$ in this lemma we have
\begin{corollary}\label{cor:apr30_5}
The following relation holds: $L(\lambda,{a}) =
M(\lambda,\underline{a})$.
\end{corollary}
We give an example for Lemma \ref{lem:may27_9}.
%%%%%%%%%%%%%%%%%%%%%%%%%%%%%%%%%%%%%%%%%%%%%%%
\begin{example}
Recall the first rigged configuration
of the  example in Section~\ref{sec:1}.
\begin{displaymath}
\setlength{\unitlength}{5mm}
\begin{picture}(8,4)(0,0.5)
\youngdg
\put(1,0){\makebox(1,1){$6$}}
\put(1,1){\makebox(1,1){$8$}}
\put(2,2){\makebox(1,1){$0$}}
\put(5.1,3){\makebox(1,1){$-5$}}
\put(-3,1.5){\makebox(2,1){$(\lambda,\underline{a})=$}}
\end{picture}
\end{displaymath}
The set of riggings is
$\underline{a} = \underline{a}^1 \sqcup 
\underline{a}^2 \sqcup \underline{a}^3$ with
$\underline{a}^1 = \{ 6,8 \}, 
\underline{a}^2 = \{ 0 \}, \underline{a}^3 = \{ -5 \}$.
%%%%%%%%%
We have
(by denoting $M_i(\lambda,\underline{a})$ by $M_i$)
\begin{align*}
&\Phi_1 \circ \Omega_{\underline{a}^3}(M_4) = {-4 \choose -3},
\quad
\Phi_1^2 \circ \Omega_{\underline{a}^3}(M_4) = {-3 \choose -1},
\quad
M_3 = \Phi_1^3 \circ \Omega_{\underline{a}^3}(M_4) = 
{-2 \choose 1},\\
&M_2 = \Phi_1 \circ \Omega_{\underline{a}^2}(M_3) =
\left(
\begin{array}{cl}
-2 & \textit{0} \\
\textit{0} & 1 
\end{array}
\right) + 
\left(
\begin{array}{ll}
1 & 3 \\
2 & 4 
\end{array}
\right) =
\left(
\begin{array}{cl}
-1 & 3 \\
2 & 5 
\end{array}
\right), 
\\
&M_1 = \Phi_1 \circ \Omega_{\underline{a}^1}(M_2) =
\left(
\begin{array}{clll}
-1 & 3 & \textit{6} & \textit{8}  \\
2 & 5 & \textit{6} & \textit{8}
\end{array}
\right) +
\left(
\begin{array}{llll}
1 & 3 & 5 & 7 \\
2 & 4 & 6 & 8
\end{array}
\right) =
\left(
\begin{array}{llll}
0 & 6 & 11 & 15 \\
4 & 9 & 12 & 16 
\end{array}
\right).
\end{align*}
Compare the $M_3,M_2,M_1$ in this example
with the $L_3,L_2,L_1$ in Example \ref{ex:may27_6}.
\end{example}
\vspace{1cm}
%%%%%%%%%%%%%%%%%%%%%%%%%%%%%%%%%%%%%%%%%%%%%%%
%
Now we present the main result of this paper.
By combining Corollaries \ref{cor:apr30_4} with
\ref{cor:apr30_5} we have
\begin{theorem}\label{th:apr30_6}
If the state of the box-ball system associated with
$M(\lambda,\underline{a})$ is updated once, it is
the state of the system associated with
$M(\lambda,{a})$.
\end{theorem}
%%%%%%%%%%%%%%%%%%%%%%%%%%%%%
By this theorem we can regard the rows of $\lambda$ as the action 
variables for the box-ball system, and the
riggings as the associated angle variables.
Initial value problems of the box-ball system 
can now be solved by an {\em inverse scattering method}.
Let $T(a)$ be the rigging made out of $a$
by replacing $a^i_j$ by $a^i_j + 
\sum_{k=1}^i h_k$.
Let $T(M) \in \Hz$ be the matrix for the
automaton state which is made out of the state 
for $M \in \Hz$ by updating once.
Then for any integer $N$ we have
%%%%%%%%%%%%%%%%%%%%%%%
\begin{displaymath}
\begin{CD}
\begin{bmatrix}
\text{Automaton State} \\
M \in \Hz 
\end{bmatrix}
@>\text{Direct Scattering}>> 
\begin{bmatrix}
\text{Scattering Data} \\
(\lambda,a) \in \R 
\end{bmatrix}
\\
@V\text{Time Evolution}VV 
@VV\text{Linearized Time Evolution}V \\
\begin{bmatrix}
\text{Automaton State} \\
T^N(M) \in \Hz 
\end{bmatrix}
@<\text{Inverse Scattering}<< 
\begin{bmatrix}
\text{Scattering Data} \\
(\lambda,T^N(a)) \in \R
\end{bmatrix}.
\end{CD}
\end{displaymath}
This diagram shows that
the time evolution of any automaton state is
given by a composition of the direct scattering,
the linearized time evolution of the scattering data, and
the inverse scattering.

%%%%%%%%%%%%%%%%%%%%%%%%%%%%%%%%%%%%%%%%%%%%%%%%%%
To close the section
we recall the notion of soliton content in
Definition~\ref{def:may26_2}.
We can write the soliton content $\lambda$ as
a partition
\begin{equation}\label{eq:may25_2}
\lambda =(1^{m_1} 2^{m_2} \ldots h^{m_h}).
\end{equation}
Here $m_i \in \Zn$ is the multiplicity of $i$ in $\lambda$.
We assume $m_h \ne 0$;
if we take $\lambda$ as in Figure \ref{fig:1}
then $h = \sum_{k=1}^l h_k$, namely the width of 
the Young diagram.
We can say that
there are $m_h$ solitons of length $h$,
$m_{h-1}$ solitons of length $h-1$ , ... , and
$m_{1}$ solitons of length $1$ in the automaton state
associated with $\lambda$.
The numbers of
successive balls in the {real space} can change.
However the solitons (in the space of scattering data)
keep their independence.
To see the soliton content in the real space one 
generally has to apply the updating process many times.
%%%%%%%%%%%%%%%%%%%%%%%%%%%%%%%%%%%%%%%%%%%%%%%%%%%%%%%%%%
\section{Upper and lower bounds for the 
inverse scattering transform}
\label{sec:6}
Before describing the aim of this section, here
we give a representation for rigged configurations that is
slightly different from the previous one.
Let $\lambda$ be the Young diagram of the form (\ref{eq:may25_2}).
Denote the set of riggings for $i$ by
$J^i = \{ J^i_j \}_{1 \leq j \leq m_i}, \,
J^i_j \in \Z$.
In other words $J^i$ is associated with the set of 
$\lambda$'s rows of length $i$.
Then if $m_i=0$ we have $J^i=\emptyset$.
Let $J = \sqcup_{1 \leq i \leq h} J^i$.
Now the pair $(\lambda,J)$ specifies a rigged configuration.

\begin{remark}
For reasons of simpler descriptions of some statements
we shall assume
$J^i_1 \leq J^i_2 \leq \cdots \leq J^i_{m_i}$
for any $i$.
\end{remark}

In this section we study
the upper and lower bounds for the
inverse scattering transform.
The result is used in the following sections to
study the relation between
partition functions for the box-ball system
and fermionic formulas.
% 25 May
It is also used to
prove that our inverse scattering transform
is equivalent to
a bijection in \cite{KKR} (Section~\ref{sec:9}).

For any set of integers $X$
we denote $\Phi_1 \circ \Omega_X$ by $\Pi_X$, where
$\Phi_1$ and $\Omega_X$ are the maps introduced in
Section~\ref{sec:4}.
We set
\begin{align}
\M_a(\lambda,J) &= \Pi_{J^a} \circ \Pi_{J^{a+1}} \circ 
\cdots \circ \Pi_{J^h} (\M_{h+1}),\label{eq:may12_3}\\
\M(\lambda,J) &= \M_1(\lambda,J),\label{eq:may13_1}
\end{align}
where $\M_{h+1} \in \Hz_0$ is the matrix with no
entry.
We can regard $\M$ as a mapping from $\R$ to $\Hz$.
It should be observed that
this $\M$ is the same as the $M$  in Section~\ref{sec:5}
although the descriptions of rigged configurations
are different.
By abuse of notation we shall denote $\varXi^{-1}$
either the $M$ or the $\M$.

We try to find expressions for the lower right and the
upper left elements of the matrices $\M_a(\lambda,J)$ in
(\ref{eq:may12_3}) because they determine the upper
and lower bounds for the inverse scattering
transform.
Given any $i$ with $m_i > 0$
let $c^{i,a} = \sum_{k=a}^h 
\left( \min(k,i) - a + 1 \right) m_k$
and $J^{i,a}_j = J^i_j + 2 c^{i,a}$ for
$1 \leq a \leq i+1$.
It is easy to see that 
\begin{equation}\label{eq:apr30_8}
J^{a-1,a}_j = J^{a-1}_j
\end{equation}
and
\begin{equation}\label{eq:apr30_7}
c^{i,a+1} = c^{i,a} - (m_a + \cdots + m_h).
\end{equation}
For $1 \leq a \leq h$ we have
\begin{lemma}\label{lem:may7_2}
\par\noindent
\begin{enumerate}
\item The lower right element of
$\Omega_{J^a} (\M_{a+1}(\lambda,J))$ is
$\max_{a \leq i \leq h, m_i>0} \{ J^{i,a+1}_{m_i} \}$.
\item The lower right element of
$\M_{a}(\lambda,J)$ is
$\max_{a \leq i \leq h,m_i>0} \{ J^{i,a}_{m_i} \}$.
\end{enumerate}
\end{lemma}
\begin{proof}
We give a proof by descending induction on $a$.
For $a=h$ the item $1$ holds because both sides
are equal to $J^h_{m_h}$.
Assume that the item $1$ holds for $a$.
Since $\M_a(\lambda,J) = \Phi_1 \circ \Omega_{J^a}
(\M_{a+1}(\lambda,J))$ we have
\begin{align*}
\mbox{L.~R.~E. of} \,
\M_{a}(\lambda,J) &= 
\mbox{L.~R.~E. of} \,
\Omega_{J^a} (\M_{a+1}(\lambda,J)) +
2(m_a + \cdots + m_h) \\
&= \max_{a \leq i \leq h,m_i>0} \{ J^{i,a+1}_{m_i} \}
+ 2(m_a + \cdots + m_h) \\
&= \max_{a \leq i \leq h,m_i>0} \{ J^{i,a}_{m_i} \}.
\end{align*}
(L.~R.~E. is for {\em lower right element}.)
Here the first equality is by the definition of $\Phi_1$,
the second is by the assumption of induction, and the
third is from (\ref{eq:apr30_7}).
Therefore the item $2$ holds for $a$.
Then we have
\begin{align*}
\mbox{L.~R.~E. of} \,
\Omega_{J^{a-1}} (\M_{a}(\lambda,J))
&= \max\{ J^{a-1}_{m_{a-1}}, 
\max_{a \leq i \leq h,m_i>0} \{ J^{i,a}_{m_i} \} \} \\
&= \max_{a-1 \leq i \leq h,m_i>0} \{ J^{i,a}_{m_i} \}.
\end{align*}
Here we used (\ref{eq:apr30_8}).
Therefore the item $1$ holds for $a-1$.
The proof is completed.
\end{proof}
%
% \begin{corollary}\label{coro:may12_1}
% The lower right element of
% $\varXi^{-1}(\lambda,J)$ is
% $\max_{1 \leq i \leq h,m_i>0} \{ J^i_{m_i} +
% 2 \sum_{j=1}^h \min(i,j) m_j\}$.
% \end{corollary}

For $1 \leq a \leq h$ we have
\begin{lemma}\label{lem:may10_1}
\par\noindent
\begin{enumerate}
\item The upper left element of
$\Omega_{J^a} (\M_{a+1}(\lambda,J))$ is
$\min_{a \leq i \leq h, m_i>0} \{ J^i_1 + i-a \}$.
\item The upper left element of
$\M_{a}(\lambda,J)$ is
$\min_{a \leq i \leq h,m_i>0} \{ J^i_1 + i-a +1\}$.
\end{enumerate}
\end{lemma}
\begin{proof}
We give a proof by descending induction on $a$.
For $a=h$ the item $1$ holds because both sides
are equal to $J^h_{m_h}$.
Assume that the item $1$ holds for $a$.
By the definition of $\Phi_1$ the item $2$ for $a$
follows.
Then we have
\begin{align*}
\mbox{U.~L.~E. of} \,
\Omega_{J^{a-1}} (\M_{a}(\lambda,J))
&= \min\{ J^{a-1}_1, 
\min_{a \leq i \leq h,m_i>0} \{ J^i_1 + i-a +1\} \} \\
&= \min_{a-1 \leq i \leq h,m_i>0} \{ J^i_1 + i - (a-1)\}.
\end{align*}
(U.~L.~E. is for {\em upper left element}.)
Therefore the item $1$ holds for $a-1$.
The proof is completed.
\end{proof}
%
% \begin{corollary}\label{coro:may12_2}
% The upper left element of
% $\varXi^{-1}(\lambda,J)$ is
% $\min_{1 \leq i \leq h, m_i>0} \{ J^i_1 +i \}$.
% \end{corollary}
%%%%%%%%%%%%%%%%%%%%%%%%%%%%%%%%%%%%%%%
Then we have that
\begin{theorem}\label{th:may13_6}
Let $(\lambda,J)$ be a rigged configuration of the
form given at the beginning of this section.
For the automaton state associated with
$\varXi^{-1}(\lambda,J)$ the following statements hold.
\par\noindent
\begin{enumerate}
\item The number of balls is
$| \lambda | := \sum_{h \geq i \geq 1} i m_i$.
\item The left-most wave tail of the state is
$\min_{1 \leq i \leq h, m_i>0} \{ J^i_1 +i \}$,
and the right-most wave front is
$\max_{1 \leq i \leq h,m_i>0} \{ J^i_{m_i} +
2 \sum_{j=1}^h \min(i,j) m_j\}$.
\end{enumerate}
\end{theorem}
\begin{proof}
Item 1 follows from the fact that
each of the $\Pi_{J^a}$ in (\ref{eq:may12_3}) is embedding
as many $\empfil$s as the height of the $a$-th
column of the Young diagram $\lambda$, which is equal to
$m_a + \cdots + m_h$.
Item 2 follows from Lemmas
\ref{lem:may7_2} and \ref{lem:may10_1}.
\end{proof}
%%%%%%%%%%%%%%%%%%%%%%%%%%%%%%%%%%%%%%%%%%%%%%%%%%%%%
%%%%%%%%%%%%%%%%%%%%%%%%%%%%%%%%%%%%%%%%%%%%%%%
\section{Automaton states in a finite interval}
\label{sec:7}
In the set of all automaton states $\Hz$ we define
a set of all the states, and
a set of all the {\em highest weight}
states in a finite interval with a specified soliton content.
Then we determine the associated subsets in the
set of all rigged configurations $\R$.
The results are used in the next section.

To begin with we introduce the notion of
\begin{definition}[highest weight state]
\label{def:may14_3}
A state of the box-ball system is called a
{\em highest weight state} if all the balls are
on the right-hand side of the wall $0$, and
for any $\ell$ the number of
empty boxes between the walls $0$ and $\ell$
is not less than that of filled boxes there.
\end{definition}
Consider an array of $L$ boxes.
We put the numbers $0,1,\ldots,L$ to its walls.
Recall the matrix
notation in Section~\ref{sec:3}.
Let $\H_n (L,\lambda)$ be the set of all matrices
of the form (\ref{eq:apr23_1}) obeying
the conditions
$0 \leq \alpha_1, \alpha_{2n} \leq L$ and
$\xi(M)=\lambda$,
and $\H_n^+ (L,\lambda)$ be its subset 
with additional conditions
$2 \sum_{k=1}^{j-1}\alpha_{2k} + \alpha_{2j} \leq
2 \sum_{k=1}^{j}\alpha_{2k-1}$ for $1 \leq j \leq n$.
%25 May
In the latter case
the set of additional conditions is equivalent to 
that for the highest weight states in
Definition \ref{def:may14_3}.
We set
\begin{align}
\H (L,\lambda) &= \sqcup_{n \geq 0} \H_n (L,\lambda),
\label{eq:may27_2}\\
\H^+ (L,\lambda) &= \sqcup_{n \geq 0} \H_n^+ (L,\lambda).
\label{eq:may27_3}
\end{align}
They are sets of automaton states in the finite interval
between walls $0$ and $L$,
and with the specified soliton content $\lambda$.

Recall the rigged configuration $(\lambda,J)$ of the form
given at the beginning of Section~\ref{sec:6}.
Fix a positive integer $L$.
As functions of the variable $\lambda$ we write
\begin{align}
p_i(\lambda) :&=L - 2 \sum_{j \geq 1} \min(i,j) m_j, 
\label{eq:may28_2}\\
m_i(\lambda):&=m_i, \quad
\phi (\lambda):=\sum_{i,j \geq 1} \min(i,j) m_i m_j.\label{eq:may28_3}
\end{align}
The latter will be used in the next section
(Lemma~\ref{lem:may28_4}).
We set
\begin{align}
\R (L,\lambda) &= \{ (\mu,J) \in \R |
\mu = \lambda, -i \leq J^i_1 \leq \cdots \leq J^i_{m_i} \leq
p_i(\lambda) \, \mbox{for any $i$} \}, 
\label{eq:may27_4} \\
\R^+ (L,\lambda) &= \{ (\mu,J) \in \R |
\mu = \lambda, 0 \leq J^i_1 \leq \cdots \leq J^i_{m_i} \leq
p_i(\lambda) \, \mbox{for any $i$} \}.
\label{eq:may27_5}
\end{align}

By definition we have
$\H^+(L,\lambda) \subset \H(L,\lambda)$ and 
$\R^+(L,\lambda) \subset \R(L,\lambda)$.
{}From Theorem \ref{th:may13_6} it is clear that 
\begin{lemma}\label{lem:may27_1}
The inverse scattering transform is a bijection
between  $\H(L,\lambda)$ and
$\R(L,\lambda)$.
\end{lemma}
Moreover we have that
\begin{lemma}\label{lem:may14_5}
The inverse scattering transform is a bijection
between  $\H^+(L,\lambda)$ and
$\R^+(L,\lambda)$.
\end{lemma}
\begin{proof}
Recall that
the map $\varXi^{-1}$ is so defined as to make automaton states
by embedding $\empfil$s recursively
at the wave tails, and at the wall
positions specified by the riggings.
We assume $J^i_{m_i} \leq p_i(\lambda)$ for any $i$;
if otherwise
the right-most wave front of $\varXi^{-1}(\lambda,J)$
would exceed $L$.
Suppose $(\lambda,J) \in \R^+(L,\lambda)$.
By induction we prove the automaton state associated with
$\M_a(\lambda,J)$ in (\ref{eq:may12_3}) is 
a highest weight state.
If $a=h+1$ the claim holds by definition.
Suppose the automaton state associated with 
$\M_{a+1}(\lambda,J)$ is a highest weight state.
By the embeddings
we are shifting the divided arrays of boxes to
the right (See Example \ref{ex:may14_4}).
Therefore if a state is a highest weight state, then the
state obtained from it by embedding a $\empfil$ at
any wave tail, or at any non-negative wall position
is a highest weight state.
Thus by the assumption of induction
the automaton state associated with $\M_{a}(\lambda,J)$ 
is also a highest weight state.
By taking $a=1$ we have
$\varXi^{-1}(\lambda,J) \in \H^+(L,\lambda)$.
Thus
$\varXi^{-1}(\R^+(L,\lambda)) \subset \H^+(L,\lambda)$.

Conversely suppose $(\lambda,J) \notin \R^+(L,\lambda)$.
Then there is a negative rigging $J^i_1(<0)$ for some $i$.
By embedding a $\empfil$ at $J^i_1$ we obtain a
state that is not a highest weight state.
One cannot obtain a highest weight state from a
non-highest weight state by embedding 
a $\empfil$ at any position.
Therefore we have $\varXi^{-1}(\lambda,J) \notin 
\H^+(L,\lambda)$.
Thus $\varXi(\H^+(L,\lambda)) \subset \R^+(L,\lambda)$.
\end{proof}
%%%%%%%%%%%%%%%%%%%%%%%%%%%%%%%%%%%%%%%%%%%%%%%%%%%%%%%%%%%%%%%
\section{Partition functions with a specified soliton content}
\label{sec:8}
In this section we study partition functions
for the box-ball system.
Recall the quantity $m_i$ in (\ref{eq:may25_2}).
In the context of the box-ball system
the $m_i$ was the number of {\em solitons} of length $i$.
By considering the partition function taken over the highest
weight states, we shall find that the $m_i$ 
coincides with a well-known
quantity in the context of Bethe ansatz, the number of
{\em strings} of length $i$.
We also consider a partition function taken over not
necessarily highest weight states.

First we introduce the notion of
energy for states of the box-ball
system.
Among the
four types of configurations of adjacent boxes, let
$\empemp,\,\filfil,\,\filemp$ do not cost the energy,
and let $\empfil$ cost the energy of 
its center wall position.
\begin{definition}[energy of an automaton state]
\label{def:may12_5}
For any matrix $M \in \Hz$
of the form (\ref{eq:apr23_1})
(or for the automaton state associated with $M$),
the quantity
$E_{\rm CTM}(M):=
\sum_{k \geq 1} \alpha_{2k-1}$ is called
its {\em energy}.
\end{definition}
This energy is not a conserved quantity of the box-ball system.
It has its origin in the corner transfer matrix
(CTM) analyses of the
solvable lattice models \cite{B}.
In particular
the energy in Definition \ref{def:may12_5}
is for the six-vertex model
\cite{KMN} but the sign was reversed 
according to the change of the model to its 
ferromagnetic regime.

Next
we introduce the notion of 
\begin{definition}[energy of a rigged configuration]
\label{def:may12_4}
For any rigged configuration $(\lambda,J)$
of the form
at the beginning of Section~\ref{sec:6}
the quantity
$E_{\rm RC}(\lambda,J):=
\sum_{h \geq i,j \geq 1} \min(i,j) m_i m_j +
\sum_{i=1}^h \sum_{j=1}^{m_i} J^i_j$ is called
its {\em energy}.
\end{definition}

We observe that
the inverse scattering transform 
is an energy-preserving bijection in the sense of these
energies.
\begin{lemma}\label{lem:may12_6}
If $\varXi^{-1}(\lambda,J) = M$
then the identity $E_{\rm RC}(\lambda,J)=
E_{\rm CTM}(M)$ holds.
\end{lemma}
\begin{proof}
Fill $1$'s, $3$'s, $5$'s, $\ldots$ into the boxes of the
first, second, third, $\ldots$ rows of 
the Young diagram $\lambda$.
The sum of these numbers in the $a$-th column is
$(\sum_{a=k}^h m_a)^2$ because the length of
the column is $(\sum_{a=k}^h m_a)$.
By summing them for $1 \leq a \leq h$ we obtain
$\sum_{h \geq i,j \geq 1} \min(i,j) m_i m_j$.
Then by adding the sum of all the riggings we have
$E_{\rm RC}(\lambda,J)$.
Let $M=\varXi^{-1}(\lambda,J)$.
Recall
the map $\Pi_{J^a}=\Phi_1 \circ \Omega_{J^a}$ in
(\ref{eq:may12_3}) and consider what it does on
the first row of $\M_{a+1}(\lambda,J)$.
The $\Omega_{J^a}$ embeds $J^a_1,\ldots,J^a_{m_a}$ 
somewhere and the $\Phi_1$ adds $1,3,5,\ldots$ to
the first, second, third, $\ldots$ columns.
By summing them for $1 \leq a \leq h$ we obtain
$E_{\rm CTM}(M)$ that is certainly equal to
the above $E_{\rm RC}(\lambda,J)$.
\end{proof}
%%%%%%%%%%%%%%%%%%%%%%%%%%%%%%%%%%%%%%%%%%%%
Let us introduce the notion of partition functions for
the box-ball system with a specified soliton
content \cite{KOTY}.
By means of
the sets of automaton states (\ref{eq:may27_2}),
(\ref{eq:may27_3}) and
the energy in Definition \ref{def:may12_5}
we define
\begin{align}
Z(L,\lambda) &:=\sum_{M \in \H (L,\lambda)} q^{E_{\rm CTM}(M)},\\
Z^+(L,\lambda) &:=\sum_{M \in \H^+ (L,\lambda)} q^{E_{\rm CTM}(M)}.
\end{align}
%
%%%%%%%%%%%%%%%%%%%%%%%%%%%%%%%%%%%%%%%%%%
On the other hand
we define the partition functions
for the rigged configurations
by using (\ref{eq:may27_4}), (\ref{eq:may27_5})
and the energy in
Definition \ref{def:may12_4} as
\begin{align}
Y(L,\lambda) &:= \sum_{(\lambda,J) \in \R (L,\lambda)} 
q^{E_{\rm RC}(\lambda,J)},\\
Y^+(L,\lambda) &:= \sum_{(\lambda,J) \in \R^+ (L,\lambda)} 
q^{E_{\rm RC}(\lambda,J)}.
\end{align}
%%%%%%%%%%%%%%%%%%%%%%%%%%
Recall the definition of the
$q$-binomial coefficient \cite{A}
\begin{displaymath}
{ n \brack m} = 
\begin{cases}
\frac{\prod_{k=1}^n (1-q^k)}
{\prod_{k=1}^m (1-q^k) \prod_{k=1}^{n-m} (1-q^k)}
& \text{if  $0 \leq m \leq n$},\\
0 & \text{otherwise.}
\end{cases}
\end{displaymath}
%%%%%%%%%%%%%%%%%%%%%%%%%%
Then we have that
\begin{lemma}\label{lem:may28_4}
The following relations hold,
\begin{align*}
Y(L,\lambda) &= 
%  \sum_{\stackrel{\lambda}{|\lambda| =s}} 
q^{-|\lambda|+\phi(\lambda)}
\prod_{i \geq 1} { p_i(\lambda) + m_i(\lambda) + i 
\brack  m_i(\lambda) },\\
Y^+(L,\lambda) &= 
%  \sum_{\stackrel{\lambda}{|\lambda| =s}} 
q^{\phi(\lambda)}
\prod_{i \geq 1} { p_i(\lambda) + m_i(\lambda) 
\brack  m_i(\lambda) }.
\end{align*}
Here $p_i(\lambda), m_i(\lambda)$ and $\phi(\lambda)$
are given by (\ref{eq:may28_2}),(\ref{eq:may28_3}).
\end{lemma}
\begin{proof}
The latter relation follows from the
identity of the $q$-binomial coefficient \cite{A}
\begin{displaymath}
\sum_{0 \leq J_1 \leq \cdots \leq J_m \leq p}
q^{J_1+\cdots+J_m} =
{ p+m \brack m }.
\end{displaymath}
The former relation also follows from this identity
by replacing $p$ by $p+i$ and $J_j$'s by $J_j+i$'s.
\end{proof}
By Lemmas \ref{lem:may27_1},
\ref{lem:may14_5} and \ref{lem:may12_6} 
the identities $Z(L,\lambda)=Y(L,\lambda)$ and 
$Z^+(L,\lambda)=Y^+(L,\lambda)$ hold.
Thus we have that
\begin{theorem}\label{prop:may26_3}
The following relations hold,
\begin{align*}
\sum_{M \in \H (L,\lambda)} q^{E_{\rm CTM}(M)} &= 
q^{-|\lambda|+\phi(\lambda)}
\prod_{i \geq 1} { p_i(\lambda) + m_i(\lambda) + i 
\brack  m_i(\lambda) },\\
\sum_{M \in \H^+ (L,\lambda)} q^{E_{\rm CTM}(M)} &= 
q^{\phi(\lambda)}
\prod_{i \geq 1} { p_i(\lambda) + m_i(\lambda) 
\brack  m_i(\lambda) }.
\end{align*}
\end{theorem}
%%%%%%%%%%%%%%%%%%%%%%%%%%%%%%%%%%%%%%%%%%%
These are the identities proposed in \cite{KOTY}.

By taking a sum over all the soliton contents
with a fixed number $(s=|\lambda|)$ of balls
their left-hand sides yield
simple expressions (See \ref{app:a}).
Then we have that
%%%%%%%%%%%%%%%%%%%%%%%%%%%%%%%%%%
\begin{proposition}
The following relations hold.
For $0 \leq s \leq L$ 
\begin{displaymath}
{ L \brack s } = q^{-s}
% \sum_{\stackrel{\lambda}{|\lambda| =s}} q^{\phi(\lambda)}
\sum_{\lambda \vdash s} q^{\phi(\lambda)}
\prod_{i \geq 1} { p_i(\lambda) + m_i(\lambda) + i 
\brack  m_i(\lambda) },
\end{displaymath}
and for $0 \leq s \leq L/2$
\begin{displaymath}
{ L \brack s } - { L \brack s-1} =
\sum_{\lambda \vdash s} q^{\phi(\lambda)}
\prod_{i \geq 1} { p_i(\lambda) + m_i(\lambda) 
\brack  m_i(\lambda) }.
\end{displaymath}
\end{proposition}
The latter is a
$q$-analogue of the Bethe's formula \cite{Be}.
In the context of Bethe ansatz
the summation variable
$\lambda =(1^{m_1} 2^{m_2} \ldots)$ 
has meant that the associated eigenstate of the Heisenberg magnet
is given by a set of variables which satisfy
an algebraic equation (the Bethe equation), where
the multiplicity of 
the strings (particular configurations of
its roots in the complex plane)
of length $i$ was given by $m_i$.
Thus we obtained a new interpretation of a
well-known fermionic character formula;
{\em the summation variable $\lambda$ is representing the
soliton content of a cellular automaton.}

On the other hand the former formula may be new and
the author expects that it can also be used in
a completeness problem of Bethe ansatz analysis for some 
quantum mechanical system.
%%%%%%%%%%%%%%%%%%%%%%%%%%%%%%%%%%%%%%%%%%%%%%%%%%%%%%%%%%%%%%%
\section{An alternative description of the
inverse scattering transform}\label{sec:9}
In this section we describe 
$sl(2)$ case of the bijection in \cite{KKR}
in our setting
and prove that it is equivalent to 
our inverse scattering transform.

%%%%%%%%%%%%%%%%%%%%%%%%%%%%%%%%%%%%%%%%%%%%%%%%%%%
Let $(\lambda,J)$ be the rigged configuration
given at the beginning of Section~\ref{sec:6}.
We show a procedure to
make another rigged configuration $(\lambda',J')$ from it.
Let $s$ be the smallest integer with $m_s>0$
that maximizes
$J^{s,1}_{m_s} = J^s_{m_s} + 2 \sum_{j=1}^h \min(s,j) m_j$.
Call the row of the diagram $\lambda$ of length $s$ and
with rigging $J^s_{m_s}$ the 
{\em shortest singular row} \cite{KKR}. 
We first delete the right-most box 
of the shortest singular row, 
and then rearrange the order of
the rows of the diagram
so that the result is again a Young diagram.
We replace the rigging of the
shortest singular row by a specific manner; explicitly
\begin{enumerate}
\item If $s=1$ let $\lambda' =(1^{m_1-1} 2^{m_2}
\ldots h^{m_h})$, $J'^1=J^1 \setminus \{ J^1_{m_1}\}$, and
$J'^i =J^i$ for $2 \leq i \leq h$.
\item If $s \geq 2$ let $\lambda' =(1^{m_1}
\ldots (s-1)^{m_{s-1}+1} s^{m_s-1} \ldots
h^{m_h})$, 
$J'^{s-1} = J^{s-1} \sqcup \{ J^{s,s}_{m_s} -1 \}$,
$J'^s=J^s \setminus \{ J^s_{m_s}\}$, and
$J'^i =J^i$ for $i \ne s,s-1$.
\end{enumerate}
Here
$J^{s,s}_{m_s} = J^s_{m_s}+2(m_s + \cdots + m_h)$.
\begin{theorem}
If $s=1$ the matrix $\M(\lambda',J')$ is made out of
$\M(\lambda,J)$ by
removing its right-most column.
If $s \geq 2$ it is made out of $\M(\lambda,J)$
by subtracting $1$ from its lower right element.
\end{theorem}
\begin{proof}
For both cases we have
$\M_a(\lambda',J') = \M_a(\lambda,J)$ for 
$s+1 \leq a \leq h$.
Since $J'^s=J^s \setminus \{ J^s_{m_s}\}$ we see that
the matrix 
$\Omega_{J'^s} (\M_{s+1}(\lambda',J') ) $ is made out of
$\Omega_{J^s} (\M_{s+1}(\lambda,J) ) $ by removing
its right-most column $^t\!(J^s_{m_s},J^s_{m_s})$.
By applying $\Phi_1$ we can deduce that
$\M_{s}(\lambda',J')  $ is made out of
$\M_{s}(\lambda,J)  $ by removing
its right-most column 
$^t\!(J^{s,s}_{m_s}-1,J^{s,s}_{m_s})$.
Hence the $s=1$ case follows.
Suppose $s \geq 2$.
Then since 
$J'^{s-1} = J^{s-1} \sqcup \{ J^{s,s}_{m_s} -1 \}$
we see that
$\Omega_{J'^{s-1}} (\M_{s}(\lambda',J') ) $ and
$\Omega_{J^{s-1}} (\M_{s}(\lambda,J) ) $ have the same
number of columns;
we also see that
the right-most column of the former is 
$^t\!(J^{s,s}_{m_s}-1,J^{s,s}_{m_s}-1)$,
that of the latter is
$^t\!(J^{s,s}_{m_s}-1,J^{s,s}_{m_s})$, and
the other columns are equal to each other.
By applying $\Phi_1$ again we obtain
$\M_{s-1}(\lambda',J') $ with its lower right element
$J^{s,s-1}_{m_s}-1$, and
$\M_{s-1}(\lambda,J) $ with its lower right element
$J^{s,s-1}_{m_s}$.
By the proof of Lemma \ref{lem:may7_2} one can deduce that
their difference is kept untouched
until we obtain 
$\M(\lambda',J')$ with its lower right element
$J^{s,1}_{m_s}-1$, and
$\M(\lambda,J)$ with its lower right element
$J^{s,1}_{m_s}$.
\end{proof}
{}From this theorem and its proof we have that
\begin{corollary}
The automaton state $\varXi^{-1}(\lambda',J')$ is made out of
$\varXi^{-1}(\lambda,J)$ by
removing its right-most ball.
\end{corollary}
This implies that
our inverse scattering transform 
for the box-ball system
coincides with the bijection in \cite{KKR}.
In the $sl(2)$ case the latter method 
(in our terminology) determines the positions of the balls
by repeating the procedure for obtaining $(\lambda',J')$ from
$(\lambda,J)$.
We note that although their method was defined only for
the highest weight states, it can be also defined for
the non-highest weight states in the $sl(2)$ case.
%%%%%%%%%%%%%%%%%%%%%%%%%%%%%%%%%%%%%%%%%%%%%%%%%%%%%%%%%%%%%%%

\vspace{0.4cm}
\noindent
\textbf{Acknowledgements} \hspace{0.1cm}
The author thanks Atsuo Kuniba, Masato Okado,
and Yasuhiko Yamada for valuable discussions
and a collaboration in the previous work where he was able 
to learn the idea of the inverse scattering method.
%%%%%%%%%%%%%%%%%%%%%%%%%%%%%%%%%%%%%%%%%%%%%%%%%%%%%%
% \appendix
\setcounter{equation}{0}
\setcounter{section}{0}
\renewcommand{\thesection}{Appendix \Alph{section}}
\renewcommand{\theequation}{A \arabic{equation}}
\section{Partition functions without a specified
soliton content}\label{app:a}
The following 
items are included here to make this paper to be self-contained,
although they can be found elsewhere.

\begin{figure}[htbp]
%WinTpicVersion3.08
\unitlength 0.1in
\begin{picture}( 37.4000, 24.2000)(0,-26.0000)
% LINE 2 0 3 0
% 28 1200 400 1200 2600 1200 400 4400 400 1200 1000 1400 1000 1400 1000 1400 600 1400 600 1600 600 1600 600 1600 400 1200 1400 1400 1400 1400 1200 1400 1200 1600 1200 1600 1200 1600 800 1600 800 1800 800 1800 800 1800 600 1800 600 2200 600 2200 600 2200 400 2200 400
% 
\special{pn 8}%
\special{pa 1200 400}%
\special{pa 1200 2600}%
\special{fp}%
\special{pa 1200 400}%
\special{pa 4400 400}%
\special{fp}%
\special{pa 1200 1000}%
\special{pa 1400 1000}%
\special{fp}%
\special{pa 1400 1000}%
\special{pa 1400 600}%
\special{fp}%
\special{pa 1400 600}%
\special{pa 1600 600}%
\special{fp}%
\special{pa 1600 600}%
\special{pa 1600 400}%
\special{fp}%
\special{pa 1200 1400}%
\special{pa 1400 1400}%
\special{fp}%
\special{pa 1400 1200}%
\special{pa 1400 1200}%
\special{fp}%
\special{pa 1600 1200}%
\special{pa 1600 1200}%
\special{fp}%
\special{pa 1600 800}%
\special{pa 1600 800}%
\special{fp}%
\special{pa 1800 800}%
\special{pa 1800 800}%
\special{fp}%
\special{pa 1800 600}%
\special{pa 1800 600}%
\special{fp}%
\special{pa 2200 600}%
\special{pa 2200 600}%
\special{fp}%
\special{pa 2200 400}%
\special{pa 2200 400}%
\special{fp}%
% LINE 2 0 3 0
% 14 1400 1400 1400 1200 1400 1200 1600 1200 1600 1200 1600 800 1600 800 1800 800 1800 800 1800 600 1800 600 2200 600 2200 600 2200 400
% 
\special{pn 8}%
\special{pa 1400 1400}%
\special{pa 1400 1200}%
\special{fp}%
\special{pa 1400 1200}%
\special{pa 1600 1200}%
\special{fp}%
\special{pa 1600 1200}%
\special{pa 1600 800}%
\special{fp}%
\special{pa 1600 800}%
\special{pa 1800 800}%
\special{fp}%
\special{pa 1800 800}%
\special{pa 1800 600}%
\special{fp}%
\special{pa 1800 600}%
\special{pa 2200 600}%
\special{fp}%
\special{pa 2200 600}%
\special{pa 2200 400}%
\special{fp}%
% LINE 2 0 3 0
% 24 1200 1800 1400 1800 1400 1800 1400 1600 1400 1600 1600 1600 1600 1600 1600 1400 1600 1400 1800 1400 1800 1400 1800 1000 1800 1000 2000 1000 2000 1000 2000 800 2000 800 2400 800 2400 800 2400 600 2400 600 3200 600 3200 600 3200 400
% 
\special{pn 8}%
\special{pa 1200 1800}%
\special{pa 1400 1800}%
\special{fp}%
\special{pa 1400 1800}%
\special{pa 1400 1600}%
\special{fp}%
\special{pa 1400 1600}%
\special{pa 1600 1600}%
\special{fp}%
\special{pa 1600 1600}%
\special{pa 1600 1400}%
\special{fp}%
\special{pa 1600 1400}%
\special{pa 1800 1400}%
\special{fp}%
\special{pa 1800 1400}%
\special{pa 1800 1000}%
\special{fp}%
\special{pa 1800 1000}%
\special{pa 2000 1000}%
\special{fp}%
\special{pa 2000 1000}%
\special{pa 2000 800}%
\special{fp}%
\special{pa 2000 800}%
\special{pa 2400 800}%
\special{fp}%
\special{pa 2400 800}%
\special{pa 2400 600}%
\special{fp}%
\special{pa 2400 600}%
\special{pa 3200 600}%
\special{fp}%
\special{pa 3200 600}%
\special{pa 3200 400}%
\special{fp}%
% LINE 2 0 3 0
% 34 1200 2200 1400 2200 1400 2200 1400 2000 1400 2000 1600 2000 1600 2000 1600 1800 1600 1800 1800 1800 1800 1800 1800 1600 1800 1600 2000 1600 2000 1600 2000 1200 2000 1200 2200 1200 2200 1200 2200 1000 2200 1000 2600 1000 2600 1000 2600 800 2600 800 3400 800 3400 800 3400 600 3400 600 3600 600 3600 600 3600 400 3600 400 3600 400
% 
\special{pn 8}%
\special{pa 1200 2200}%
\special{pa 1400 2200}%
\special{fp}%
\special{pa 1400 2200}%
\special{pa 1400 2000}%
\special{fp}%
\special{pa 1400 2000}%
\special{pa 1600 2000}%
\special{fp}%
\special{pa 1600 2000}%
\special{pa 1600 1800}%
\special{fp}%
\special{pa 1600 1800}%
\special{pa 1800 1800}%
\special{fp}%
\special{pa 1800 1800}%
\special{pa 1800 1600}%
\special{fp}%
\special{pa 1800 1600}%
\special{pa 2000 1600}%
\special{fp}%
\special{pa 2000 1600}%
\special{pa 2000 1200}%
\special{fp}%
\special{pa 2000 1200}%
\special{pa 2200 1200}%
\special{fp}%
\special{pa 2200 1200}%
\special{pa 2200 1000}%
\special{fp}%
\special{pa 2200 1000}%
\special{pa 2600 1000}%
\special{fp}%
\special{pa 2600 1000}%
\special{pa 2600 800}%
\special{fp}%
\special{pa 2600 800}%
\special{pa 3400 800}%
\special{fp}%
\special{pa 3400 800}%
\special{pa 3400 600}%
\special{fp}%
\special{pa 3400 600}%
\special{pa 3600 600}%
\special{fp}%
\special{pa 3600 600}%
\special{pa 3600 400}%
\special{fp}%
\special{pa 3600 400}%
\special{pa 3600 400}%
\special{fp}%
% LINE 2 0 3 0
% 38 1200 2400 1200 2400 1600 2400 1600 2400 1600 2200 1600 2200 1800 2200 1800 2200 1800 2000 1800 2000 2000 2000 2000 2000 2000 1800 2000 1800 2200 1800 2200 1800 2200 1400 2200 1400 2400 1400 2400 1400 2400 1200 2400 1200 2800 1200 2800 1200 2800 1000 2800 1000 3600 1000 3600 1000 3600 800 3600 800 3800 800 3800 800 3800 600 3800 600 4200 600 4200 600 4200 400 4200 400
% 
\special{pn 8}%
\special{pa 1200 2400}%
\special{pa 1200 2400}%
\special{fp}%
\special{pa 1600 2400}%
\special{pa 1600 2400}%
\special{fp}%
\special{pa 1600 2200}%
\special{pa 1600 2200}%
\special{fp}%
\special{pa 1800 2200}%
\special{pa 1800 2200}%
\special{fp}%
\special{pa 1800 2000}%
\special{pa 1800 2000}%
\special{fp}%
\special{pa 2000 2000}%
\special{pa 2000 2000}%
\special{fp}%
\special{pa 2000 1800}%
\special{pa 2000 1800}%
\special{fp}%
\special{pa 2200 1800}%
\special{pa 2200 1800}%
\special{fp}%
\special{pa 2200 1400}%
\special{pa 2200 1400}%
\special{fp}%
\special{pa 2400 1400}%
\special{pa 2400 1400}%
\special{fp}%
\special{pa 2400 1200}%
\special{pa 2400 1200}%
\special{fp}%
\special{pa 2800 1200}%
\special{pa 2800 1200}%
\special{fp}%
\special{pa 2800 1000}%
\special{pa 2800 1000}%
\special{fp}%
\special{pa 3600 1000}%
\special{pa 3600 1000}%
\special{fp}%
\special{pa 3600 800}%
\special{pa 3600 800}%
\special{fp}%
\special{pa 3800 800}%
\special{pa 3800 800}%
\special{fp}%
\special{pa 3800 600}%
\special{pa 3800 600}%
\special{fp}%
\special{pa 4200 600}%
\special{pa 4200 600}%
\special{fp}%
\special{pa 4200 400}%
\special{pa 4200 400}%
\special{fp}%
% LINE 0 0 3 0
% 36 1200 2400 1600 2400 1600 2400 1600 2200 1600 2200 1800 2200 1800 2200 1800 2000 1800 2000 2000 2000 2000 2000 2000 1800 2000 1800 2200 1800 2200 1800 2200 1400 2200 1400 2400 1400 2400 1400 2400 1200 2400 1200 2800 1200 2800 1200 2800 1000 2800 1000 3600 1000 3600 1000 3600 800 3600 800 3800 800 3800 800 3800 600 3800 600 4200 600 4200 600 4200 400
% 
\special{pn 20}%
\special{pa 1200 2400}%
\special{pa 1600 2400}%
\special{fp}%
\special{pa 1600 2400}%
\special{pa 1600 2200}%
\special{fp}%
\special{pa 1600 2200}%
\special{pa 1800 2200}%
\special{fp}%
\special{pa 1800 2200}%
\special{pa 1800 2000}%
\special{fp}%
\special{pa 1800 2000}%
\special{pa 2000 2000}%
\special{fp}%
\special{pa 2000 2000}%
\special{pa 2000 1800}%
\special{fp}%
\special{pa 2000 1800}%
\special{pa 2200 1800}%
\special{fp}%
\special{pa 2200 1800}%
\special{pa 2200 1400}%
\special{fp}%
\special{pa 2200 1400}%
\special{pa 2400 1400}%
\special{fp}%
\special{pa 2400 1400}%
\special{pa 2400 1200}%
\special{fp}%
\special{pa 2400 1200}%
\special{pa 2800 1200}%
\special{fp}%
\special{pa 2800 1200}%
\special{pa 2800 1000}%
\special{fp}%
\special{pa 2800 1000}%
\special{pa 3600 1000}%
\special{fp}%
\special{pa 3600 1000}%
\special{pa 3600 800}%
\special{fp}%
\special{pa 3600 800}%
\special{pa 3800 800}%
\special{fp}%
\special{pa 3800 800}%
\special{pa 3800 600}%
\special{fp}%
\special{pa 3800 600}%
\special{pa 4200 600}%
\special{fp}%
\special{pa 4200 600}%
\special{pa 4200 400}%
\special{fp}%
% LINE 2 1 3 0
% 4 1200 2600 4400 2600 4400 2600 4400 400
% 
\special{pn 8}%
\special{pa 1200 2600}%
\special{pa 4400 2600}%
\special{da 0.070}%
\special{pa 4400 2600}%
\special{pa 4400 400}%
\special{da 0.070}%
% STR 2 0 3 0
% 3 1270 250 1270 350 2 0
% $\beta_1$
\put(12.7000,-3.5000){\makebox(0,0)[lb]{$\beta_1$}}%
% STR 2 0 3 0
% 3 1770 260 1770 360 2 0
% $\beta_2$
\put(17.7000,-3.6000){\makebox(0,0)[lb]{$\beta_2$}}%
% STR 2 0 3 0
% 3 2580 260 2580 360 2 0
% $\beta_3$
\put(25.8000,-3.6000){\makebox(0,0)[lb]{$\beta_3$}}%
% STR 2 0 3 0
% 3 3710 250 3710 350 2 0
% $\beta_{p-1}$
\put(37.1000,-3.5000){\makebox(0,0)[lb]{$\beta_{p-1}$}}%
% STR 2 0 3 0
% 3 4270 260 4270 360 2 0
% $\beta_{p}$
\put(42.7000,-3.6000){\makebox(0,0)[lb]{$\beta_{p}$}}%
% STR 2 0 3 0
% 3 3260 250 3260 350 2 0
% $\cdots$
\put(32.6000,-3.5000){\makebox(0,0)[lb]{$\cdots$}}%
% STR 2 0 3 0
% 3 680 700 680 800 2 0
% $\alpha_1+1$
\put(6.8000,-8.0000){\makebox(0,0)[lb]{$\alpha_1+1$}}%
% STR 2 0 3 0
% 3 890 1190 890 1290 2 0
% $\alpha_2$
\put(8.9000,-12.9000){\makebox(0,0)[lb]{$\alpha_2$}}%
% STR 2 0 3 0
% 3 890 1560 890 1660 2 0
% $\alpha_3$
\put(8.9000,-16.6000){\makebox(0,0)[lb]{$\alpha_3$}}%
% STR 2 0 3 0
% 3 900 1900 900 2000 2 0
% $\vdots$
\put(9.0000,-20.0000){\makebox(0,0)[lb]{$\vdots$}}%
% STR 2 0 3 0
% 3 840 2280 840 2380 2 0
% $\alpha_{p-1}$
\put(8.4000,-23.8000){\makebox(0,0)[lb]{$\alpha_{p-1}$}}%
% STR 2 0 3 0
% 3 660 2510 660 2610 2 0
% $\alpha_p-1$
\put(6.6000,-26.1000){\makebox(0,0)[lb]{$\alpha_p-1$}}%
% LINE 0 0 3 0
% 2 1200 400 1200 2400
% 
\special{pn 20}%
\special{pa 1200 400}%
\special{pa 1200 2400}%
\special{fp}%
% LINE 0 0 3 0
% 2 1200 400 4200 400
% 
\special{pn 20}%
\special{pa 1200 400}%
\special{pa 4200 400}%
\special{fp}%
\end{picture}%
\caption{A graphical representation of
the energy preserving bijection 
between automaton states to restricted partitions.
The largest Young diagram (thick line) is made of
$p-1$ border strips (skew Young diagrams without 2 by 2
square blocks).}
\label{fig:3}
\end{figure}
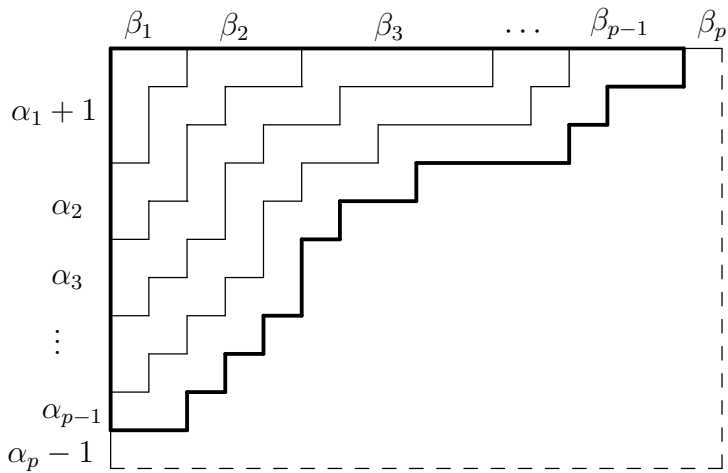

Recall the sets of the automaton states 
(\ref{eq:may27_2}), (\ref{eq:may27_3}).
We define
\begin{align*}
\H (L,s) &= \sqcup_{\lambda \vdash s} \H (L,\lambda),\\
\H^+ (L,s) &= \sqcup_{\lambda \vdash s} \H^+ (L,\lambda).
\end{align*}
(The notation is a bit ambiguous but no confusion
should occur.)
They are sets of automaton states in the finite interval
between walls $0$ and $L$,
without a specified soliton content but with the number of
balls set to be $s$.
With the energy in Definition~\ref{def:may12_5}
we define
\begin{align*}
Z(L,s) &:=\sum_{M \in \H (L,s)} q^{E_{\rm CTM}(M)},\\
Z^+(L,s) &:=\sum_{M \in \H^+ (L,s)} q^{E_{\rm CTM}(M)}.
\end{align*}
Then we have that
\begin{lemma}
The partition function $Z(L,s)$ satisfies
the recursion relation
\begin{equation} \label{eq:may13_2}
Z(L,s) = Z(L-1,s) + \sum_{k=1}^s q^{L-k}
Z(L-k-1,s-k).
\end{equation}
The $Z^+(L,s)$ also satisfies
the recursion relation of the same form.
\end{lemma}
\begin{proof}
Consider the array of $L$ boxes between walls $0$ and $L$.
The first term in the right-hand side of (\ref{eq:may13_2})
is for
those automaton states without ball
in the right-most box.
In the summation, the $k$-th term is for
those states which have their right-most empty box 
on the left of the wall $L-k$.
\end{proof}
%%%
Each of the
partition functions is uniquely determined by the 
recursion relation of the form
(\ref{eq:may13_2}) under
the boundary conditions
\begin{align}
Z(L,0)& =Z(L,L) =1,\label{eq:may13_3}\\
Z^+(L,0)&=1, Z^+(2s-1,s)=0.\label{eq:may13_4}
\end{align}
%%%%%%%%%%%%%%%%%%%%%%%%%
\begin{proposition}
The following identities hold.
\begin{align}
Z(L,s) &=
{ L \brack s }, \label{eq:may14_8}\\
Z^+(L,s) &=
{ L \brack s }- { L \brack s-1}. \label{eq:may17_1}
\end{align}
\end{proposition}
\begin{proof}
Using the identities of the $q$-binomial coefficients \cite{A}
\begin{displaymath}
{ L \brack s } = 
{ L-1 \brack s } + q^{L-s} { L-1 \brack s-1 } = 
q^s { L-1 \brack s } + { L-1 \brack s-1 },
\end{displaymath}
one can deduce that the expressions in the
right hand sides of (\ref{eq:may14_8}) and (\ref{eq:may17_1})
satisfy the
recursion relation of the form
(\ref{eq:may13_2}).
They also satisfy
the boundary conditions (\ref{eq:may13_3}) and
(\ref{eq:may13_4}) respectively.
\end{proof}
The right hand side of (\ref{eq:may14_8}) is
the generating function for the number of partitions
of an integer into at most $s$ part, and each part
is less than or equal to $L-s$ \cite{A}.
An energy preserving bijection between $\H(L,s)$ and
the set of all such partitions is given as follows.
Given $M \in \H(L,s)$ consider the automaton state associated
to $M$.
If its left-most $s$ boxes are filled,
then it is mapped to $\emptyset$.
Suppose the state is not the case.
Then the state admits the following description:
there are $\alpha_1$ filled boxes from the left,
then $\beta_1$ empty boxes, 
$\alpha_2$ filled boxes,
$\beta_2$ empty boxes, ..., 
$\alpha_p$ filled boxes,
$\beta_p$ empty boxes.
Here $p$ is an integer $\geq 2$,
$\sum_{k=1}^p \alpha_k = s$, and
$\sum_{k=1}^p \beta_k = L-s$.
We assume $\alpha_k, \beta_k \geq 1$ except
$\alpha_1, \beta_p \geq 0$.
Then the state is mapped to the largest Young diagram
in Figure \ref{fig:3}.
%%%%%%%%%%%%%%%%%%%%%%%%%%%%%%%%%%%%%%%%%%
%%%%%%%%%%%%%%%%%%%%%%%%%%%%%%%%%%%%%%%%%%%%

Conversely given an arbitrary Young diagram we can
read off these numbers $\alpha_k,\beta_k$ in
the following way.
We peel off a border strip from the south-east
direction of the Young diagram.
The result is also a Young diagram.
Therefore we can repeat this procedure as many
times as possible.
In this way any Young diagram can be divided into such
border strips uniquely.
Then we can read off the numbers $\alpha_k,\beta_k$ 
{}from the lengths of the west and north edges of
the border strips.

%%%%%%%%%%%%%%%%%%%%%%%%%%%%%%%%%%%%%%%%%%%%%%%%%%%%%%%%%%%%%

% \clearpage

%%%%%%%%%%%%%%%%%%%%%%%%%%%%%%%%%%%%%%%%%%%%%%%%%%%%%%%%%%%%%%%%

%%%%%%%%%%%%%%%%%%%%%%%%%%%%%%%%%%%%%%%%%%%%%%%%%%%%%%%%%%%

\begin{thebibliography}{99}
% \renewcommand{\baselinestretch}{2}
% \setlength{\baselineskip}{1cm}

\bibitem{T} D. Takahashi,
{On some soliton systems defined by using
boxes and balls}, Proceedings of
the International Symposium on Nonlinear Theory and
Its Applications (NOLTA '93),
(1993) 555--558.

\bibitem{TS} D. Takahashi and J. Satsuma,
{A soliton cellular automaton},
J. Phys. Soc. Jpn. {\bf 59} (1990) 3514--3519.

\bibitem{TNS} T. Tokihiro, A. Nagai and J. Satsuma,
Proof of solitonical nature of box and ball systems by
means of inverse ultra-discretization,
Inverse Probl. {\bf 15} (1999) 1639--1662.

\bibitem{TTMS} T. Tokihiro, D. Takahashi, J. Matsukidaira and J. Satsuma,
{{}From soliton equations to integrable cellular automata through
a limiting procedure},
Phys. Rev. Lett. {\bf 76} (1996) 3247--3250.

\bibitem{FOY}
K.~Fukuda, M.~Okado and Y.~Yamada, 
{Energy functions in box ball systems},
Int.\ J.\ Mod.\ Phys.\ A {\bf 15} (2000) 1379--1392.

\bibitem{TTS} M.~Torii, D.~Takahashi and J.~Satsuma,
{Combinatorial representation of invariants of a 
soliton cellular automaton},
Physica {\bf D 92} (1996) 209--220.

\bibitem{KOTY} A. Kuniba, M. Okado, T. Takagi and Y. Yamada,
{Vertex operators and partition functions for the box-ball
system}, 
Research Institute for Mathematical Sciences
(Kyoto Univ. )
K\^oky\^uroku {\bf 1302},
(2003)
91--107 [In Japanese].

\bibitem{KKR}
S.V.Kerov, A.N.Kirillov and N.Yu.Reshetikhin,
{Combinatorics, Bethe ansatz, and representations of the 
symmetric group},
Zap. Nauch. Semin. LOMI. \ {\bf 155} (1986) 50-64.


\bibitem{YYT}
D.~Yoshihara, F.~Yura and T.~Tokihiro,
Fundamental cycle of a periodic box-ball system,
J. Phys. A: Math. Gen.{\bf 36} (2003) 99--121.


\bibitem{AC}
M.J.~Ablowitz and P.A.~Clarkson,
{Solitons, Nonlinear Evolution Equations and
Inverse Scattering}, Cambridge Univ.~Press, 
(1991).
\bibitem{HIK}
K. Hikami, R. Inoue and Y. Komori,
Crystallization of the Bogoyavlensky lattice,
{J. Phys. Soc. Jpn.} {\bf  68}:  2234--2240 (1999).



\bibitem{HKT}
G. Hatayama, A. Kuniba and T.  Takagi,
{Soliton cellular automata associated with crystal bases},
Nucl. Phys. B{\bf 577}[PM] (2000) 619--645.


\bibitem{HHIKTT}
G. Hatayama, K. Hikami, R. Inoue, A. Kuniba, T. Takagi and T. Tokihiro,
{The $A^{(1)}_M$ Automata related to crystals of symmetric tensors},
J. Math. Phys. {\bf 42} (2001) 274--308.


\bibitem{HKOTY1} G.\ Hatayama,  A.\ Kuniba,
M.\ Okado, T.\ Takagi, Y.\ Yamada, 
{Remarks on fermionic formula},
Contemporary Math.\ {\bf 248} (AMS 1999) 243--291. 

\bibitem{HKOTT}
G. Hatayama, A. Kuniba, M. Okado, T. Takagi and Z. Tsuboi,
{Paths, Crystals and Fermionic Formulae},
Prog. in Math. Phys. {MathPhys Odyssey 2001, 
Integrable Models and Beyond},  M. Kashiwara  and T. Miwa eds.
Birkh{\"a}user (2002) 205--272.


\bibitem{Ma}
I.~Macdonald,
{Symmetric functions and Hall polynomials},
2nd edition, Oxford Univ. Press, New York (1995).


\bibitem{KR}
A.~N.~Kirillov and N.~Yu.~Reshetikhin,
{The Bethe ansatz and the combinatorics of Young tableaux},
J. Sov. Math. {\bf 41} (1988) 925--955.

\bibitem{NY}
A.~Nakayashiki and Y.~Yamada,
{Kostka polynomials and energy functions in solvable lattice models},
Selecta Mathematica, New Ser. {\bf 3} (1997) 547--599.


\bibitem{B}
R.J.~Baxter,
{Exactly solved models in statistical mechanics}, Academic Press, 
London (1982).

\bibitem {KMN}
S-J.~Kang, M.~Kashiwara, K.~C.~Misra, 
T.~Miwa, T.~Nakashima and A.~Nakayashiki,
{Affine crystals and vertex models},
Int.\ J.\ Mod.\ Phys.\ {\bf A7} (suppl. 1A), (1992) 449--484.

\bibitem{A}
G.E.~Andrews,
{The Theory of Partitions}, Cambridge Univ.~Press, 
(1984).

\bibitem{Be}
H.~A.~Bethe,
{Zur Theorie der Metalle, I. Eigenwerte und
Eigenfunktionen der linearen Atomkette},
Z. Physik {\bf 71} (1931) 205--231.
\end{thebibliography}
\end{document}